\documentclass[reprint,groupedaddress,superscriptaddress,aps,prd,preprintnumbers,onecolumn,11pt,fleqn,nofootinbib,eqsecnum,floatfix,showpacs,preprintnumbers,a4paper]{revtex4-2}

\usepackage{xcolor,color}
\usepackage{bm,amsmath,amssymb,dsfont}
\usepackage{graphicx}
\usepackage{afterpage}
\usepackage{enumerate}

\long\def\comment#1{ }

\newcommand{\beq}{\begin{equation}}
\newcommand{\eeq}{\end{equation}}
\newcommand{\bal}{\begin{align}}
\newcommand{\eal}{\end{align}}

\newcommand{\del}{\partial}

\newcommand{\nn}{\nonumber\\}
\newcommand{\rmd}{{\rm d}}
\newcommand{\dif}{{\rm d}}

\newcommand{\bk}{\bm{k}}

\newcommand{\bx}{\bm{x}}
\newcommand{\by}{\bm{y}}

\newcommand{\br}{\bm{r}}
\newcommand{\bb}{\bm{b}}
\newcommand{\xbj}{x_{_{\rm Bj}}}
\newcommand{\Ybj}{Y_{_{\rm Bj}}}
\newcommand{\xP}{x_{\mathbb P}}
\newcommand{\YP}{Y_{\mathbb P}}
\newcommand{\mcal}{\mathcal}
\newcommand{\bK}{\bm{K}}
\newcommand{\bP}{\bm{P}}

\begin{document}

\title{Incoherent diffractive production of jets in electron DIS off nuclei at high energy}

\author{Benjamin Rodriguez-Aguilar}
\email{benjaroagui@gmail.com}
\affiliation{European Centre for Theoretical Studies in Nuclear Physics and Related Areas (ECT*)\\
and Fondazione Bruno Kessler, Strada delle Tabarelle 286, 38123 Villazzano (TN), Italy}
\affiliation{Physics Department, Trento University, Via Sommarive 14, 38123 Povo (TN), Italy}

\author{D.N.~Triantafyllopoulos}
\email{trianta@ectstar.eu}
\affiliation{European Centre for Theoretical Studies in Nuclear Physics and Related Areas (ECT*)\\
and Fondazione Bruno Kessler, Strada delle Tabarelle 286, 38123 Villazzano (TN), Italy}

\author{S.Y.~Wei}
\email{shuyi@sdu.edu.cn}
\affiliation{Key Laboratory of Particle Physics and Particle Irradiation (MOE), Institute of frontier and interdisciplinary science, Shandong University, Qingdao, Shandong 266237, China}

\date{\today}

\begin{abstract}
We study incoherent diffractive production of two and three jets in electron-nucleus deep inelastic scattering (DIS) at small $\xbj$ using the color dipole picture and the effective theory of the Color Glass Condensate (CGC). We consider color fluctuations in the CGC weight-function as the source of the nuclear break-up and the associated momentum transfer $\sqrt{|t|}$. We focus on the regime in which the two jets are almost back-to-back in transverse space and have transverse momenta $P_{\perp}$ much larger than both the momentum transfer and the saturation scale $Q_s$. The cross section for producing such a hard dijet is parametrically dominated by large size fluctuations in the projectile wave-function that scatter strongly and for which a third, semi-hard, jet appears in the final state. The 2\,+\,1 jets cross section can be written in a factorized form in terms of incoherent quark and gluon diffractive transverse momentum distributions (DTMDs) when the third jet is explicit, or incoherent diffractive parton distribution functions (DPDFs) when the third jet is integrated over. We find that the DPDFs and the corresponding cross section saturate logarithmically when $|t| \ll Q_s^2$, while they fall like $1/|t|^2$ in the regime $Q_s^2 \ll |t| \ll P_{\perp}^2$. We further show that there is no angular correlation between the hard jet momentum and the momentum transfer. For typical EIC kinematics the 2 jets and 2\,+\,1 jets cross sections are of the same order.

\end{abstract}

\maketitle

\section{Introduction}
\label{sec:intro}

At high energy the perturbative piece of the light-cone wavefunction (LCWF) of a hadron or a nucleus consists of gluons whose occupation numbers increase rapidly if their longitudinal momentum fraction $x$ becomes very small \cite{Lipatov:1976zz,Kuraev:1977fs,Balitsky:1978ic,Mueller:1993rr}. This can cause violations of unitarity limits for various physical quantities, like when the amplitude for a projectile color dipole to scatter off the hadron or the nucleus at fixed impact parameter exceeds unity. Perturbative QCD addresses this issue by introducing non-linear dynamics that tame the growth of these highly occupied modes and predicts the phenomenon of gluon saturation  \cite{Gribov:1984tu,Mueller:1985wy}. A dynamical semi-hard scale called the saturation momentum $Q_s$ emerges, separating dilute (where gluons have transverse momenta $k_{\perp} \gg Q_s$) and dense (where $k_{\perp} \ll Q_s$) regimes. Roughly, this scale grows as $Q_s^2(A,x) \propto \Lambda^2 A^{1/3} (1/x)^\lambda$, with $\Lambda$ being the QCD scale, $A$ the atomic number, and $\lambda \simeq 0.25$ \cite{GolecBiernat:1999qd,Triantafyllopoulos:2002nz}, hence giving a justification for the use of weak coupling methods for sufficiently large nuclei and/or small-$x$. The Color Glass Condensate (CGC) \cite{Gelis:2010nm,Kovchegov:2012mbw} is a modern effective theory constructed in a way that includes gluon saturation and results in non-linear evolution equations for scattering correlators \cite{Balitsky:1995ub,Kovchegov:1999yj,JalilianMarian:1997gr,Weigert:2000gi,Iancu:2000hn,Ferreiro:2001qy}, which provide the building blocks for determining cross sections or other physical observables of interest.

The study of saturation typically involves high-energy collisions of the system under consideration with smaller projectiles. These include proton-nucleus ($pA$) collisions, Deep Inelastic Scattering (DIS) of electrons off hadrons (like $ep$) or nuclei ($eA$), and even ultra-peripheral nucleus-nucleus collisions (UPCs), where a quasi-real photon from one nucleus probes the other nucleus. During the last years, there has been significant interest in events, in which the kinematics of one or more particles in the final state are specified and the simplest such example is single inclusive DIS \cite{Iancu:2020jch,Bergabo:2022zhe,Caucal:2024cdq,Caucal:2024bae,Altinoluk:2024vgg}. A more commonly studied process is the inclusive production of two jets in the forward direction of the projectile \cite{Marquet:2007vb,Metz:2011wb,Dominguez:2011br,Mueller:2013wwa,Kotko:2015ura,Dumitru:2015gaa,Kotko:2017oxg,Dumitru:2018kuw,Klein:2019qfb,Mantysaari:2019hkq,Iancu:2020mos,Hatta:2021jcd,Boussarie:2021ybe,Caucal:2021ent,Taels:2022tza,Caucal:2023fsf}. In $ep$ and $eA$ collisions these jets originate from the $q\bar{q}$ pair fluctuation of the virtual photon, while in $pA$ collisions they usually arise from a quark-gluon pair which is formed from the splitting of a collinear quark. Very often, the focus is on the limit where the two jets are hard and almost back-to-back. Then the dijet imbalance is of the order of the momentum transferred from the target to the projectile and thus it is controlled by saturation physics. These arguments also apply to dihadron production \cite{Albacete:2010pg,Stasto:2011ru,Iancu:2013dta,Zheng:2014vka,Albacete:2018ruq,Bergabo:2022tcu}, as the perturbatively calculable part of the process does not change.

Equally significant is the diffractive production of jets or hadrons \cite{Bartels:1999tn,Altinoluk:2015dpi,Hatta:2016dxp,Salazar:2019ncp,Mantysaari:2019hkq,Iancu:2021rup,Hatta:2022lzj,Beuf:2022kyp,Iancu:2022lcw,Iancu:2023lel,Tong:2023bus,Kar:2023jkn,Shao:2024nor,Hauksson:2024bvv}. A process is considered diffractive, if there is no net color exchange between the projectile and the hadronic or nuclear target. This results in a rapidity gap, that is, a wide angular region in the final state which does not contain any particles. The target may either remain intact or break apart after the collision and we refer to these processes as coherent diffraction and incoherent diffraction, respectively. As we shall explain in Sect.~\ref{sec:general}, incoherent diffraction is proportional to fluctuations in the target \cite{Good:1960ba}. In particular, these can be color fluctuations \cite{Marquet:2010cf,Mantysaari:2019hkq,Demirci:2022wuy,Rodriguez-Aguilar:2023ihz} and they can be described and calculated in the CGC framework. Here we shall build on the expertise acquired in the coherent diffractive production of two and three jets in high energy photon-nucleus collisions  \cite{Iancu:2021rup,Iancu:2022lcw,Iancu:2023lel,Hauksson:2024bvv} in order to explore the effects of color fluctuations in the corresponding incoherent process.  
  
We have organized our work as follows. In Sect.~\ref{sec:general} we give a general discussion on incoherent diffraction in the dipole frame. In Sect.~\ref{sec:2jets} we recall the well-known result for the production of exactly two jets with arbitrary kinematics. In Sect.~\ref{sec:SIDDIS} we make a digression to see how the aligned jet configurations determine the anticipated leading-twist result at large photon virtualities in incoherent single inclusive diffractive DIS (SIDDIS) and how the incoherent quark diffractive transverse momentum distribution (DTMD) emerges. In Sect.~\ref{sec:2_hard_jets} we briefly review our main results \cite{Rodriguez-Aguilar:2023ihz} for the production of two hard jets which are almost back-to-back. In Sections~\ref{sec:2_plus_g} and \ref{sec:2_plus_q} we give the cross sections for producing 2\,+\,1 jets, more precisely two hard and back-to-back jets, which are now accompanied by a third semi-hard (gluon or quark) jet. We further see how the gluon and quark DTMDs appear as the semi-hard factors in the cross sections. In Sect.~\ref{sec:distributions} we analyse the properties of the DTMDs and, after integrating over the semi-hard jet transverse momentum, of the corresponding incoherent diffractive parton distribution functions (DPDFs). In Sect.~\ref{sec:mingap} we define the observable of interest, namely dijet production with a minimum rapidity gap, and make a direct comparison between the 2 jets and 2\,+\,1 jets contributions. In Sect.~\ref{sec:conc} we conclude, in Appendix \ref{app:Gij} we show that the gluon semi-hard factor is diagonal and in Appendix \ref{app:gDPDF} we study in full detail the gluon DPDF in limiting cases in the McLerran-Venugopalan (MV) model \cite{McLerran:1993ka,McLerran:1993ni} .

\section{Incoherent diffraction in the Dipole Frame}
\label{sec:general}

We consider electron-nucleus DIS at high energy. We will neglect the ``trivial'' electron-photon electromagnetic vertex and focus only on the interaction between the virtual photon $\gamma^*$ and the nucleus with atomic number $A$. It will be convenient to adopt a frame in which both the colliding particles are ultra-relativistic and move along the same axis. In light-cone (LC) coordinates the spacelike photon with virtuality $q^{\mu} q_{\mu}=-Q^2$ has the 4-momentum $q^{\mu} = (q^+,-Q^2/2 q^+,\bm{0}_{\perp})$, with $q^+\gg \sqrt{Q^2}$ so that it carries a large plus longitudinal momentum. Neglecting the nucleon mass, the nucleus has a 4-momentum $P^{\mu}_N=(0,P_N^-,\bm{0}_{\perp})$ per nucleon, i.e.~it carries only a minus longitudinal momentum.

In this frame the virtual photon develops partonic fluctuations whose typical lifetime is $\tau \sim 2 q^+/Q^2$. The longitudinal extent of the contracted nucleus is $L=R_A/\gamma$, with $R_A \simeq A^{1/3} R_N$ its radius and $\gamma \simeq P_N^{-} R_N$ its Lorentz factor and where $R_N$ is the nucleon radius. As usual, we define the kinematic invariant
\begin{align}
	\label{xbj}
	\xbj \equiv \frac{Q^2}{2 q \mkern-2mu \cdot \mkern-2mu P_N}
	= \frac{Q^2}{2 q^+ P_N^-}
	\equiv \frac{Q^2}{s},
\end{align}  
which stands for the fraction of the target longitudinal momentum $P_N^-$ transferred to the struck quark. Then we readily find that $\tau \gg L$ when $\xbj A^{1/3} \ll 1$, a condition which is satisfied in the problem at hand, since we will typically consider that $\xbj$ is of the order of $10^{-3}$ or smaller. Thus, the projectile partons in the LCWF of the virtual photon are long-lived and can be taken as ``frozen'' during the short period in which the scattering off the target nucleus takes place. Since the lowest order fluctuation consists of a $q\bar{q}$ pair, we refer to this description as the color dipole picture and to the specific frame that we have chosen as the dipole frame. 

We are interested in the diffractive production of forward jets, where by diffraction we mean that in the final state of the collision there is a large rapidity gap, that is, a region void of particles. In the dipole frame this will happen when the partons in the photon LCWF scatter elastically off the nucleus via the exchange of a colorless combination of partons, predominantly gluons, to which we shall occasionally refer as the Pomeron ($\mathbb P$). In turn, this implies that there must be an exchange of at least two gluons at the amplitude level (four gluons at the level of the cross section), whereas this is to be contrasted with inelastic scattering, which starts with the exchange of a single gluon in a perturbative  expansion. This is a simple and intuitive picture to have in mind, however the nucleus can be at saturation and thus generate strong color fields, which means that we have to take into account multiple scattering. The state of the target after the collision defines the type of the diffractive process and there are two possibilities: if the nucleus remains intact it is coherent, while if the nucleus breaks it is incoherent.  

In order to be able to calculate any diffractive cross section, it is more than obvious that one must have a sufficient description of the nucleus, in particular of its modes that carry a small fraction of its minus longitudinal momentum, since we require the diffractive system to move in the forward direction. These soft modes are appropriately described in the CGC framework, which provides us with a suitable weight-function representing the probability to have a certain target configuration. Then, by taking the average over all possible configurations we can compute a given observable. Hence, if $T_{\rm el}$ is the $T$-matrix for the projectile to scatter elastically off the nucleus in the direct amplitude (DA) and $\bar{T}_{\rm el}$ the corresponding one in the complex conjugate amplitude (CCA), the cross section for a diffractive process complies with 
\begin{align}
	\label{sigma_diff_gen}
	\sigma^{\scriptscriptstyle \rm D}\mkern-2mu
	\propto
	\left \langle T_{\rm el}\, \bar{T}_{\rm el} \right \rangle,
\end{align}
where we have suppressed all the dependences and integrations in the various variables. Needless to say, later on we shall give precise formulae for all cross sections of interest. But independently of the details, being proportional to the square of the elastic amplitude, generally the diffractive processes are suppressed in the dilute regime (due to color transparency) and thus they are more sensitive to contributions from configurations for which the scattering is strong. Eq.~\eqref{sigma_diff_gen} contains both the coherent and the incoherent contributions and to isolate the former, one must perform the aforementioned average at the amplitude level, i.e.
 \begin{align}
	\label{sigma_diff_coh}
	\sigma^{\scriptscriptstyle \rm D}_{\rm coh}
	\propto
	\left \langle T_{\rm el} \right \rangle
	\left \langle \bar{T}_{\rm el} \right \rangle.
\end{align}
By subtracting this coherent piece from Eq.~\eqref{sigma_diff_gen} we immediately realize that the incoherent diffractive cross section is proportional to the variance of the elastic scattering amplitude, that is
\begin{align}
	\label{sigma_diff_inc}
	\sigma^{\scriptscriptstyle \rm D}_{\rm inc}
	=
	\sigma^{\scriptscriptstyle \rm D}\mkern-2mu -
	\sigma^{\scriptscriptstyle \rm D}_{\rm coh}
	\propto
	\left \langle T_{\rm el}\, \bar{T}_{\rm el} \right \rangle
	-
	\left \langle T_{\rm el} \right \rangle
	\left \langle \bar{T}_{\rm el} \right \rangle.
\end{align} 
The physical meaning of the last expression in Eq.~\eqref{sigma_diff_inc} is that incoherent diffraction is determined by fluctuations in the target weight-function. For example, these can simply be geometrical fluctuations and elaborated models based on shapes built from ``hot-spots'' have found significant success in describing experimental results when the target is a proton \cite{Mantysaari:2016ykx,Mantysaari:2016jaz}. One can also think about particle number fluctuations, which in the context of QCD and in the kinematic regime of interest can occur via Pomeron loops \cite{Iancu:2004es,Iancu:2004iy,Iancu:2005nj,Hatta:2006hs}, although such an approach has not been pursued significantly, mostly due to technical difficulties. Here we shall base our study on color fluctuations \cite{Marquet:2010cf,Mantysaari:2019hkq,Rodriguez-Aguilar:2023ihz} which originate from the color exchange among different substructures inside the nucleus. As such, the resulting cross section is suppressed by $1/N_c^2$, with $N_c$ the number of colors, when compared with the one for the corresponding coherent or inclusive process. This mechanism, which exists in the CGC effective theory and thus involves little model dependence, has been exposed in full detail in \cite{Rodriguez-Aguilar:2023ihz}, cf.~Sect.~III there. 

An important kinematic variable in diffraction is the transverse momentum $\bm{\Delta}$ transferred from the target to the projectile system. It can originate from the spatial transverse inhomogeneity of the target, thus (in the absence of potential hot-spots) one has $\Delta_{\perp} \equiv |\bm{\Delta}| \sim 1/R_A$, which for a large nucleus like Pb or Au with $A \approx 200$ becomes $\Delta_{\perp} \approx 30\mkern2mu {\rm MeV}$, a rather small scale for our purposes. On the contrary, in incoherent diffraction the momentum transfer due to color fluctuations is set by the saturation momentum which is much larger (and perturbative), that is $\Delta_{\perp} \sim Q_s \sim 1\div 1.5\mkern2mu {\rm GeV}$. Moreover, since such fluctuations are a consequence of perturbative gluon exchanges, we expect to find a Coulomb tail at very large $\Delta_{\perp}$. Such a fall-off is significantly milder than the one due to geometrical fluctuations which is exponential.

\begin{figure}
	\begin{center}
	\includegraphics[width=0.40\textwidth]{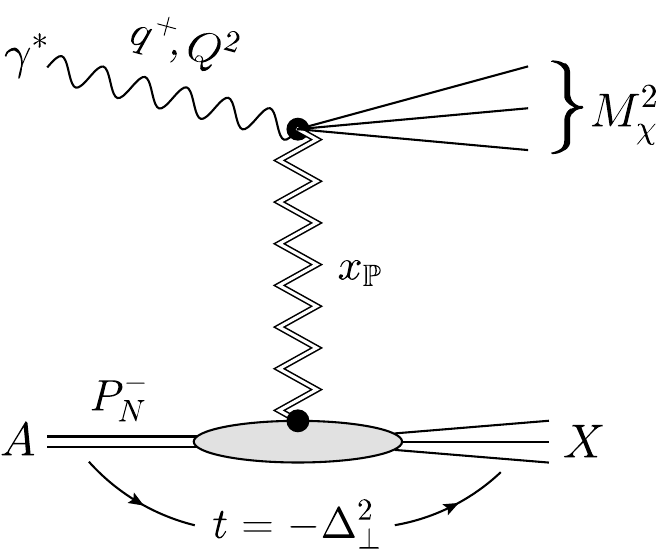}
	\end{center}
	\caption{\small Generic diagram for the incoherent reaction  $\gamma^* A \to \chi X$.}
\label{fig:diff_gen}
\end{figure}

In addition to $Q^2$ and $\xbj$, in diffraction it is customary to introduce two more kinematic invariants, namely
\begin{align}
	\label{xp_and_beta}
	x_{\mathbb P}
	\equiv
	\frac{Q^2+ M_{\chi}^2 + |t|}{s}
	\qquad \mathrm{and} \qquad
	\beta
	\equiv
	\frac{Q^2}{Q^2+ M_{\chi}^2 + |t|},
\end{align}
where $M_{\chi}^2$ is the invariant mass of the diffractive system and $t =-\Delta_{\perp}^2$ the invariant momentum transfer. In the next sections, we will explicitly determine $M_{\chi}^2$ (and therefore also $x_{\mathbb P}$ and $\beta$) in terms of the kinematics of the outgoing partonic state. Physically, $x_{\mathbb P}$ is the fraction of the target longitudinal momentum $P_N^-$ carried by the Pomeron and thus eventually transferred to the diffractive system. It should be stressed that the two variables in Eq.~\eqref{xp_and_beta} are not independent, since according to the definition of $\xbj$, cf.~\eqref{xbj}, they satisfy 
\begin{align}
	\label{xbj_xp_beta}
	\xbj = \beta x_{\mathbb P}
	\,\,\Rightarrow\,\,
	Y_{\scriptscriptstyle \rm Bj} = Y_{\beta} + Y_{\mathbb P}.
\end{align}
The first equation in the above indicates that $\beta$ is the fraction of the Pomeron longitudinal momentum transferred to the struck quark. In Eq.~\eqref{xbj_xp_beta} we have further defined the corresponding logarithmic variables, i.e.~the rapidities. $Y_{\mathbb P} \equiv \ln 1/x_{\mathbb P}$ determines the value of the desired rapidity gap, $Y_{\beta}\equiv\ln1/\beta$ corresponds to the interval occupied by the diffractive system and clearly their sum must give $Y_{\scriptscriptstyle \rm Bj}\equiv\ln 1/\xbj$. We also point out that $Q_s(x_{\mathbb P})$ is the relevant value of the saturation momentum, since the projectile as a whole probes the target at the longitudinal scale $x_{\mathbb P}$. We will typically consider the latter to satisfy $x_{\mathbb P} \lesssim 2\times10^{-2}$, which leads to gaps such that $Y_{\mathbb P} \gtrsim 4$. In Fig.~\ref{fig:diff_gen} we show the general diagram for the incoherent diffractive reaction $\gamma^* A \to \chi X$, where neither the particular dynamics nor the precise nature of the outgoing states $\chi$ and $X$ are specified.

\section{Cross Section for 2 jets}
\label{sec:2jets}

\begin{figure}
	\begin{center}
	\includegraphics[width=0.45\textwidth]{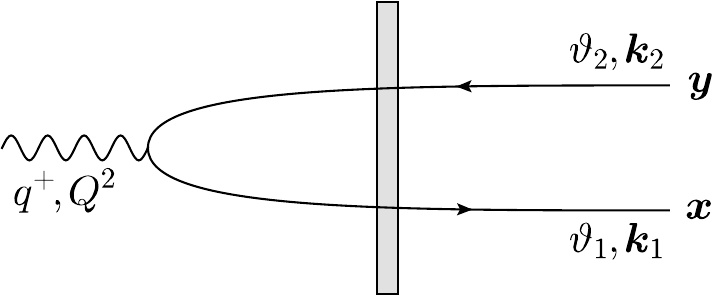}
	\end{center}
	\caption{\small Forward dijet production in DIS off a nucleus in the dipole picture. The gray vertical slab, the so-called shockwave, stands for the scattering off the nucleus. It encompasses all eikonal diagrams with an arbitrary number of gluon exchanges. In this work we assume that the total exchange, independently of how complex it may be, is always colorless and that there is a non-zero transverse momentum transfer from the nucleus to the dijet system, i.e.~$\bk_1 + \bk_2 = \bm{\Delta}$.}
\label{fig:2jets}
\end{figure}

Let us now proceed to see how the generic considerations developed in Sect.~\ref{sec:general} materialize in the simplest scenario, that in which the diffractive system is composed of exactly two jets. The virtual photon first fluctuates into a quark and an antiquark which carry corresponding fractions $\vartheta_1$ and $\vartheta_2$ of the plus longitudinal momentum $q^+$. The pair scatters elastically off the target nucleus and moreover such a scattering is  eikonal, i.e.~the aforementioned longitudinal fractions remain the same in the outgoing state. On the contrary, the transverse momenta are not only individually affected by the interaction but also the pair acquires an imbalance $\bK \equiv \bk_1 + \bk_2 = \bm{\Delta}$, where $\bk_1$ and $\bk_2$ are the final parton momenta and $\bm{\Delta}$ the transverse momentum transferred from the nucleus to the diffractive system. In Fig.~\ref{fig:2jets} we show the relevant probability amplitude.

The general expressions for dijet cross sections, either inclusive or diffractive,  have already been derived in the CGC framework and here it suffices to present the final result in the case of interest, namely \cite{Mantysaari:2019hkq,Rodriguez-Aguilar:2023ihz}
\begin{align}
\label{sigmab}
	\frac{\rmd \sigma^{\gamma_{\scriptscriptstyle T}^* 
	A\rightarrow q\bar q X}}
	{\rmd \vartheta_1 \rmd \vartheta_2\, 
	\rmd^{2} \bm{P} \rmd^{2} \bm{\Delta}}
	= \,&
	\frac{S_{\perp} \alpha_{\rm em} \,N_{c}}{4\pi^3}
	\sum\mkern-1.5mu e_f^2\,
	\delta_\vartheta
	\left(\vartheta_1^{2}+ \vartheta_2^{2} \right)
	\int \frac{\rmd^2 \bm{B}}{2\pi}\,
	\frac{\rmd^2\bm{r}}{2\pi}\,
	\frac{\rmd^2\bar{\bm{r}}}{2\pi}
	\nn
	& \times
	e^{- i \bm{\Delta} \cdot \bm{B}
	- i \bm{P} \cdot (\bm{r}-\bar{\bm{r}})}\,
	\frac{\bm{r} \!\cdot\! \bar{\bm{r}}}{r \bar{r}}\,
	\bar{Q}
	K_{1}\big(\bar{Q}r\big)
	\bar{Q}
	K_{1}\big(\bar{Q}{\bar{r}}\big)
	\mcal{W}_q(\bm{r},\bar{\bm{r}},\bm{B}),
\end{align}   
for a transversely polarized photon. To obtain the corresponding formula in the longitudinal sector it suffices to replace $ \vartheta_1^2 + \vartheta_2^2 \to 4 \vartheta_1 \vartheta_2$ and $(\br/r) K_1(\bar{Q} r) \to K_0(\bar{Q} r)$ and similarly for the respective factor in the CCA. Here $S_{\perp}\simeq \pi R_A^2$ is the transverse area of the homogeneous nucleus, $\alpha_{\rm em}$ the fine structure constant and $e_f$ the fractional electric charge of the quark with flavor $f$. We have defined the shorthand notation $\delta_\vartheta \equiv \delta(1 - \vartheta_1 - \vartheta_2)$ which stands for the conservation of the plus longitudinal momentum. In coordinate space we assume that the quark and antiquark are located at $\bx$ and $\by$ respectively. In the DA we have introduced the dipole separation $\br$ and the center of energy $\bb$ according to $\bx = \bb + \vartheta_2 \br$ and $\by=\bb - \vartheta_1\br$, while the inverse relations read $\br=\bx-\by$ and $\bb=\vartheta_1\bx+\vartheta_2 \by$ and similarly for the barred quantities in the CCA. In momentum space, along with the momentum transfer we have also defined the relative momentum via $\bk_1 = \bP +\vartheta_1 \bm{\Delta}$ and $\bk_2 = -\bP +\vartheta_2 \bm{\Delta}$, so that $\bP = \vartheta_2\bk_1 - \vartheta_1 \bk_2$. The factor $\vartheta_1^2 + \vartheta_2^2$ arises from the  $\gamma^* \to q\bar{q}$ splitting function, whereas the presence of the Bessel function $K_1(\bar{Q}r)$ (or $K_0(\bar{Q}r)$ in the longitudinal case), with $\bar{Q}^2 = \vartheta_1\vartheta_2Q^2$, is the mathematical manifestation of the fact that the parton lifetimes must be at most of the order of $2q^+/Q^2$, otherwise their emission is heavily suppressed. 
Finally, 
\begin{align}
	\label{wfund}
	\hspace{-0.4cm}
	\mcal{W}_q
	(\bm{r},\bar{\bm{r}},\bm{B}) = 
	\left \langle 
	T(\bm{x},\bm{y}) T(\bar{\bm{y}},\bar{\bm{x}}) 
	\right \rangle
	\!-\!
	\left \langle 
	T(\bm{x},\bm{y}) 
	\right \rangle \!
	\left \langle
	T(\bar{\bm{y}},\bar{\bm{x}})
	\right \rangle
	=
	\left \langle 
	S(\bm{x},\bm{y}) S(\bar{\bm{y}},\bar{\bm{x}}) 
	\right \rangle
	\!-\!
	\left \langle 
	S(\bm{x},\bm{y}) 
	\right \rangle \!
	\left \langle
	S(\bar{\bm{y}},\bar{\bm{x}})
	\right \rangle
\end{align} 
is the particular correlator which isolates the relevant QCD dynamics for incoherent diffractive scattering in the CGC effective theory. Since the target nucleus is homogeneous, translation invariance requires that any correlator depends only on the difference between any pair of transverse coordinates. Thus, in the problem at hand, $\mcal{W}_q$ depends on three independent vectors which we have conveniently chosen to be the dipole separations $\br$ and $\bar{\br}$ as well as the difference between the impact parameters in the DA and the CCA, that is $\bm{B}=\bm{b}-\bar{\bm{b}}$. In Eq.~\eqref{wfund} $T=1-S$, where $S(\bx,\by)$ is the $S$-matrix for the $q{\bar{q}}$ dipole ($\bx,\by$) to scatter elastically of the nucleus and reads
\begin{align}
	\label{s_and_V}
	S(\bx,\by) = \frac{1}{N_c}\,
	{\rm tr} \big[V(\bx) V^\dagger(\by)\big]
	\qquad \mathrm{with} \qquad
	V(\bm{x})\equiv {\rm T}
	\exp\left[i g \int \rmd x^{+} 
	t^{a} A^{-}_a(x^{+},\bm{x})\right].
\end{align} 
In the definition of the Wilson line $V(\bx)$, T indicates time ordering in the light-cone time $x^+$, $g$ is the QCD coupling, $t^a$ is a SU($N_c$) color matrix in the fundamental representation and $A^{-}_a$, with $a=1, \dots, N_c^2-1$, is the single non-vanishing component of the target color field in the Lorentz gauge $\del_{\mu} A_a^{\mu}=0$. Notice that $A^{-}_a$ does not depend on $x^-$ and therefore, as already mentioned, it cannot alter the plus longitudinal momenta of the projectile partons. The averages in Eq.~\eqref{wfund} must be performed with the CGC weight-function at the appropriate scale which is the rapidity gap $Y_{\mathbb P} = \ln 1/x_{\mathbb P}$. Typically one assumes an initial condition given by the MV model at $Y_0=\ln1/x_0$, with $x_0 \simeq 10^{-2}$, and then evolves for a rapidity interval $Y_{\mathbb P}-Y_0$. It must be recalled that, for given $Q^2$ and $\xbj$, the precise value of both $\xP$ and $\beta$ depend on the kinematics of the outgoing projectile particles. Indeed, requiring that the final quark and antiquark are individually on-shell, we recover Eq.~\eqref{xp_and_beta} with $M_{\chi}^2$ being their total invariant mass, that is 
\begin{align}
	\label{Mchi_2jets}
	M_{\chi}^2 = (k_1 + k_2)^2 = \frac{P_{\perp}^2}{\vartheta_1\vartheta_2}
	\qquad 
	\mathrm{for \mkern6mu 2 \mkern6mu jets}.
\end{align}

\section{SIDDIS, aligned jet configurations and incoherent quark DTMD}
\label{sec:SIDDIS}

We temporarily digress from our main goal to see under what circumstances the dipole picture, which follows the process from the projectile viewpoint, leads to the leading-twist result at high $Q^2$ as a partonic description of the target would suggest. In order to do so, we shall consider incoherent SIDDIS, in which process one measures the transverse momentum of a single particle in the final state at fixed $\beta$ and $|t|$. In the case under consideration (with only two particles in the final state), the longitudinal and transverse momenta of the unobserved particle are not independent variables and therefore the only thing we have to do is to change variables from, say, $\vartheta_1$ to $\beta$. From now on, we shall further assume that $P_{\perp} \gg \Delta_{\perp}$, which in  turn means that $\bP$ is approximately equal to the quark transverse momentum and $-\bP$ to the antiquark one. Then $\beta$ reduces to
\begin{align}
	\label{beta_2jets}
	\beta \simeq 
	\frac{Q^2}{Q^2 + P_{\perp}^2/\vartheta_1\vartheta_2}
	=
	\frac{\bar{Q}^2}{\bar{Q^2} + P_{\perp}^2},
\end{align}
and since $\bar{Q}^2 =\vartheta_1 (1- \vartheta_1) Q^2$, Eq.~\eqref{beta_2jets} is a quadratic equation for $\vartheta_1$ which is therefore expressed in terms of the independent variables $\beta$ and $P_{\perp}$. More precisely, the solution smaller than 1/2 is
\begin{align}
	\label{theta-sol}
	\vartheta_1 = \vartheta_*(\beta,P_{\perp}) \equiv \frac{1}{2} \left(1 - 
	\sqrt{1- \frac{4 \beta}{1-\beta} \frac{P_{\perp}^2}{Q^2}} \right),	
\end{align}
while obviously there is a second solution equal to $1-\vartheta_*$ which is larger than 1/2. Since the integrand in the cross section in Eq.~\eqref{sigmab} is symmetric under the exchange $\vartheta_1 \leftrightarrow 1- \vartheta_1$, it suffices to proceed by considering only the solution in Eq.~\eqref{theta-sol} and multiplying the result by 2. We also observe that the presence of the square root implies un upper limit for the hard jet momentum with given $Q^2$ and $\beta$, or equivalently, an upper limit in the range of allowed values for $\beta$ with given $Q^2$ and $P_{\perp}^2$, namely
\begin{align}
	\label{p_beta_max}
	P_{\perp\star}^2 = \frac{1- \beta}{4\beta}\, Q^2
	\qquad \mathrm{or} \qquad 
	\beta_{\star} = \frac{Q^2}{Q^2 + 4 P_{\perp}^2}.
\end{align}
Thus, when all other kinematic variables are fixed, a large $P_{\perp}^2$ requires a small $\beta$, which is just another manifestation of the fact that the production of hard jets ``consumes'' part of the available rapidity interval.

Eq.~\eqref{beta_2jets} also implies that $\bar{Q}^2$ is independent of $Q^2$ when expressed in terms of the new variables $\beta$ and $P_{\perp}^2$. In order to emphasize such a fact and also for our later convenience, we shall use a new notation, that is
\begin{align}
	\label{Q2M2}
	\bar{Q}^2=\frac{\beta}{1-\beta}\,P_{\perp}^2 \equiv \mcal{M}^2.
\end{align}
Starting from Eq.~\eqref{sigmab}, it is a matter of straightforward algebra to write the SIDDIS transverse cross section as
\begin{align}
	\label{sigma-qq-t}
	\frac{\dif\sigma^{\gamma_{\scriptscriptstyle T}^* A\rightarrow q\bar q X}}
	{\dif^{2}\bP
  	\dif^{2}\bm{\Delta}\,
  	\dif \ln 1/\beta} =\,
  	\frac{S_{\perp} \alpha_{\rm em} N_c}{\pi^2}\,
  	\sum\mkern-1.5mu e_f^2\,
  	\frac{1}{Q^2}\,
  	\frac{1- \frac{2 \mcal{M}^2}{Q^2}}
  	{\sqrt{1- \frac{4 \mcal{M}^2}{Q^2}}}\,
  	\frac{\mcal{Q}_{\rm inc}}{1-\beta},
  	\end{align}
 where
 \begin{align}
 	\label{Qinc}
 	\mcal{Q}_{\rm inc}\equiv\frac{1}{2\pi}
 	\int \frac{\rmd^2 \bm{B}}{2\pi}\,
	\frac{\rmd^2\bm{r}}{2\pi}\,
	\frac{\rmd^2\bar{\bm{r}}}{2\pi}
	e^{- i \bm{\Delta} \cdot \bm{B}
	- i \bm{P} \cdot (\bm{r}-\bar{\bm{r}})}\,
	\frac{\bm{r} \!\cdot\! \bar{\bm{r}}}{r \bar{r}}\,
	\mcal{M}^2
	K_{1}(\mcal{M}r)
	\mcal{M}^2
	K_{1}(\mcal{M}{\bar{r}})
	\mcal{W}_q(\bm{r},\bar{\bm{r}},\bm{B}).
\end{align}
Since $\mcal{Q}_{\rm inc}$ is independent of $Q^2$, the cross section is just proportional to $1/Q^2$ if and only if $\mcal{M}^2 \ll Q^2$, or equivalently $P_{\perp}^2 \ll P_{\perp\star}^2$. This means that the solution in Eq.~\eqref{theta-sol} automatically selects asymmetric pairs with $\vartheta_*\simeq \mcal{M}^2/Q^2 \ll 1$. In other words, the above exercise demonstrates that the leading twist result is recovered when the final state is restricted to aligned jet configurations: the fermion with fraction $1-\vartheta_*$ carries almost all of the longitudinal momentum $q^+$ of the virtual photon. 

In this $\mcal{M}^2 \ll Q^2$ limit, we shall further rewrite the cross section as
	\begin{align}
	\label{sigma-qq-dtmd}
	\frac{\dif\sigma^{\gamma_{\scriptscriptstyle T}^* A\rightarrow q\bar q X}}
	{\dif^{2}\bP
  	\dif^{2}\bm{\Delta}\,
  	\dif \ln 1/\beta} =\,
  	\frac{4 \pi^2\alpha_{\rm em}}{Q^2}\,
  	\sum\mkern-1.5mu e_f^2\,
  	2\,
  	 \frac{\rmd xq_{\mathbb{P}}^{\rm inc}
	(x, x_{\mathbb{P}}, \bP, \bm{\Delta})}
	{\rmd^2 \bP \rmd^2\bm{\Delta}}\bigg |_{x=\beta},	  	
  	\end{align}
which exhibits TMD factorization and where we have introduced the (anti)quark incoherent DTMD
\begin{align}
  \label{qDTMD}
  \frac{\rmd xq_{\mathbb{P}}^{\rm inc}
  (x, x_{\mathbb{P}}, \bP, \bm{\Delta})}
  {\rmd^2 \bP \rmd^2\bm{\Delta}}
  \equiv
  \frac{S_{\perp} N_c}{4 \pi^3}\,	
	\frac{\mcal{Q}_{\rm inc}(x, x_{\mathbb{P}}, \bP, \bm{\Delta})}
	{2\pi(1-x)}.
\end{align}
$\mcal{Q}_{\rm inc}$ is given by Eq.~\eqref{Qinc} and is viewed as a function of the rapidity related variables $x=\beta$ and $x_{\mathbb P}$ and the transverse momenta $\bP$ and $\bm{\Delta}$. More precisely, it depends on $P_{\perp}$ and $\Delta_{\perp}$, and it can also depend on  the scalar product $\bP \cdot \bm{\Delta}$, but often only $|t| = \Delta_{\perp}^2$ is measured, so that any such angular dependence disappears. Here it becomes important to stress that when we integrate over $\bm{\Delta}$, Eqs.~\eqref{sigma-qq-dtmd} and \eqref{qDTMD} become completely analogous to those for coherent diffraction with the only difference between the two cases arising from the QCD correlator. In the present study, as already repeatedly emphasized, it is given by the connected piece defined in Eq.~\eqref{wfund}. In coherent diffraction one must simply replace $\mcal{W}_q$ in Eq.~\eqref{Qinc} with the correlator $\left \langle T(\br) 
	\right \rangle \left \langle T(\bar{\br}) \right \rangle$. Then, for a homogeneous nucleus, the integrations in the DA and the CCA  factorize, the integration over $\bm{B}$ gives a $\delta$-function in $\bm{\Delta}$ and, after integration over $\bm{\Delta}$, we recover the (anti)quark DTMD in \cite{Hauksson:2024bvv}.
	
Without performing any new calculation, we shall now claim that the above result is in agreement with the following picture in the standard Bjorken frame and in the LC gauge $A_a^-=0$, in which one would normally construct the LCWF of the left-moving target. The Pomeron is viewed as a colorless partonic fluctuation of the target with a small (minus) longitudinal fraction $\xP$ and a transverse momentum $\bm{\Delta}$ and splits into a $q\bar{q}$ pair. One of the fermions, say the quark, carries a splitting fraction $1-x$   
	and transverse momentum $\bP$ and appears in the final state. The other fermion carrying a fraction $x$ and transverse momentum $-\bP +\bm{\Delta}$, with $\Delta_{\perp} \ll P_{\perp}$,\footnote{Unlike in the coherent case where $\Delta_{\perp}$ is a non-perturbative scale, here it can be of the order of $Q_s$ or even larger.} is initially virtual and is put on-shell only after it absorbs the virtual photon. As detailed in \cite{Hauksson:2024bvv}, cf.~Eqs.~(2.37)-(2.41) there, this description is correct if and only if $\vartheta_1 \ll 1$. Moreover, in this limit, the fraction $x$ is identified with the diffractive variable $\beta$ in \eqref{beta_2jets} and this explains a posteriori the use of the variable $x$ already in Eq.~\eqref{sigma-qq-dtmd}. 
	
Thus, to conclude, the above TMD factorized formula and the physical explanation in terms of partons emerging from the Pomeron require the outgoing state to be in the aligned jets configuration. 

\section{Review of the Cross Section for 2 hard jets}
\label{sec:2_hard_jets}

\begin{figure}
	\begin{center}
	\includegraphics[width=0.65\textwidth]{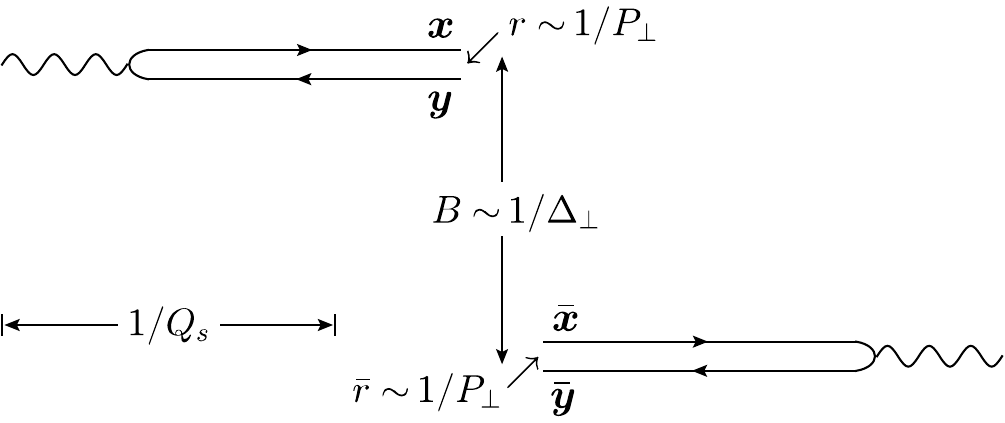}
	\end{center}
	\caption{\small A representative configuration of dipoles which contributes to the incoherent diffractive dijet cross section in Eq.~\eqref{sigmab} in the kinematic regime  $P_{\perp} \gg \Delta_{\perp}, Q_s$. The $q\bar{q}$ dipole with size  $r \sim 1/P_{\perp} \ll 1/Q_s$ in the DA is separated from the one with size  $\bar{r} \sim 1/P_{\perp} \ll 1/Q_s$ in the CCA by a large distance $B \sim 1/\Delta_{\perp}$. In the typical case $\Delta_{\perp} \sim Q_s$, the distance $B$ is of the order of $1/Q_s$.}
\label{fig:sizes_qq}
\end{figure}

Now we return to Eq.~\eqref{sigmab} to see how it simplifies when the two jets are hard and almost back-to-back, more precisely we will assume that $P_{\perp} \gg \Delta_{\perp},Q_s$. Since $\bP$ and $\bm{\Delta}$ are the momentum variables conjugate to the coordinates $\br$ (and $\bar{\br}$) and $\bm{B}$ respectively, we can expand the correlator $\mcal{W}_q(\br,\bar{\br},\bm{B})$ for $r,\bar{r} \ll B,1/Q_s$. In Fig.~\ref{fig:sizes_qq} we give a pictorial representation of such a configuration in coordinate space. Using the Gaussian approximation \cite{Iancu:2002aq,Dumitru:2011vk,Iancu:2011ns,Iancu:2011nj,Alvioli:2012ba}, this calculation was already performed in \cite{Rodriguez-Aguilar:2023ihz} and the dijet cross section was expressed as a sum of four terms, each of which is a product between a $P_{\perp}$-dependent hard factor that includes the $\gamma^*\to q\bar{q}$ splitting and a $\Delta_{\perp}$-dependent semi-hard factor that involves the dipole-nucleus scattering amplitude. Even though the final result depends explicitly on the angle between $\bP$ and $\bm{\Delta}$, here we shall be interested only in the averaged over such an angle cross section. One finds\footnote{The approximate equality stands for the fact that we neglect the term proportional to $\mcal{G}_{\rm D}^{(-)}$ in Eq.(4.13) in \cite{Rodriguez-Aguilar:2023ihz}, since it is both parametrically and numerically small.}
\begin{align}
	\label{sigmaave}
	\frac{\rmd\sigma^{\gamma_\lambda^* 
	A\rightarrow q\bar q X}}
	{\dif \Pi} \simeq  
	\frac{S_{\perp} \alpha_{\rm em} \,N_{c}}{4\pi^3}
	\sum\mkern-1.5mu e_f^2\,
	\delta_\vartheta\,
	\frac{2 C_F}{N_c^3}\,
	H_{\lambda}(\vartheta_1,\vartheta_2,P_{\perp},\bar{Q})\,
	\mcal{G}^{(+)}(\Delta_{\perp}),
\end{align}
with $C_F=(N_c^2-1)/2N_c$ and  where the shorthand notation $\dif \Pi = (2\pi)^2 P_{\perp} \dif P_{\perp} \Delta_{\perp} \dif \Delta_{\perp} \dif \vartheta_1 \dif \vartheta_2$ has been introduced. The color factor $2 C_F/N_c^3$ appears because the leading in $N_c$ term cancels when we consider the connected QCD correlator in Eq.~\eqref{wfund}. The index $\lambda$ stands for transverse ($T$) or longitudinal ($L$) polarization and the corresponding hard factors read
\begin{align}
	\label{hard_factor_2jets_T}
	& H_{T}(\vartheta_1,\vartheta_2,P_{\perp},\bar{Q}) =
	\big(\vartheta_1^2+\vartheta_2^2\big)\,
	\frac{P_{\perp}^2 (3 \bar{Q}^4 + P_{\perp}^4)}
	{(P_{\perp}^2 + \bar{Q}^2)^6},
	\\
	\label{hard_factor_2jets_L}
	& H_{L}(\vartheta_1,\vartheta_2,P_{\perp},\bar{Q}) =
	\vartheta_1\vartheta_2\,
	\frac{2 \bar{Q}^2 (\bar{Q}^4  - 2 \bar{Q}^2 P_{\perp}^2 + 5P_{\perp}^4 )}
	{(P_{\perp}^2 + \bar{Q}^2)^6}.
\end{align}
Finally, the dimensionful semi-hard factor is determined by
	\begin{align}
	\label{Gdplus}
	\mcal{G}^{(+)}(\Delta_{\perp}) = 
	\int_0^{\infty} \!\rmd B B J_0(\Delta_{\perp} B)
	\Phi(B)
	[F_+(B)]^2,
\end{align}
where the two scalar functions $\Phi$ and $F_+$ are given by
\begin{align}
	\label{Phi_and_Fplus}
	\Phi(B) = \frac{\mcal{S}_g(B) - 1 - 
	\ln \mcal{S}_g(B) }
	{\ln^2\mcal{S}_g(B)}
	\qquad \mathrm{and} \qquad
	F_+(B)= \nabla^2 \ln \mcal{S}_g(B),
\end{align}
with $\mcal{S}_g(B)$ the averaged over the target configurations $S$-matrix for the scattering of a gluon-gluon dipole, i.e.~a dipole in the adjoint representation.

There is no restriction on the values of $\vartheta_1, \vartheta_2$ and $Q^2$ for Eq.~\eqref{sigmaave} to be valid and in particular one is even allowed to take the limit $Q^2 \to 0$. Still, in order to have a well-defined problem as the one in studied in \cite{Rodriguez-Aguilar:2023ihz}, we shall assume that $\vartheta_1$ and $\vartheta_2$ are of the order of one-half and we shall generally think that $P_{\perp}^2$ and $\bar{Q}^2$ are of the same order. This means that the two hard factors in Eqs.~\eqref{hard_factor_2jets_T} and \eqref{hard_factor_2jets_L} behave like $1/P_{\perp}^6$. The origin of such a power can be understood by inspecting the starting expression in Eq.~\eqref{sigmab}: the $\gamma^*$ decay vertex is responsible for a factor $1/\bar{Q}^2 \sim 1/P_{\perp}^2$, while the expansion of $\mcal{W}_q(\br,\bar{\br},\bm{B})$ at small $r$ and $\bar{r}$ starts with $r^2 \bar{r}^2 \sim 1/P_{\perp}^4$, i.e.~the $q\bar{q}$ dipole interacts weakly with the target nucleus.

The evaluation of the semi-hard factor $\mcal{G}^{(+)}(\Delta_{\perp})$ requires the knowledge of $\mcal{S}_g(B)$ and for a generic $\Delta_{\perp}$ we must proceed numerically. However, we can gain very good qualitative understanding by assuming that the dipole-nucleus scattering is described by the MV model \cite{McLerran:1993ka,McLerran:1993ni}, that is  
\begin{align}
        \label{SgMV}
        \mcal{S}_g(B) =  1 - \mcal{T}_g(B) =
        \exp \left(  
        -\frac{B^2 Q_{A}^2}{4} 
        \ln \frac{4}{B^2\Lambda^2} \right),
\end{align}
where $\Lambda$ is a non-perturbative color neutralization scale, which can be taken as the QCD scale, and $Q_A^2$ is related to the gluon saturation momentum via 
\begin{align}
	\label{QAQs}
	Q_A^2 \ln \frac{Q_s^2}{\Lambda^2} = Q_s^2.
\end{align}  
Then one finds the piecewise expression
\begin{align}
	\label{Gplus_cases}
	\mcal{G}^{(+)}(\Delta_{\perp}) \simeq 
	\begin{cases}
	{\displaystyle Q_A^2 \left( \ln^2 \frac{Q_s^2}{\Lambda^2} - \ln^2 \frac{\Delta_{\perp}^2}{\Lambda^2} \right)}
	 \quad &\mathrm{for} \quad 
	 \Delta_{\perp} \ll Q_s,
	 \\*[0.4cm]
	 \mcal{O}\big(Q_s^2\big) 
	 \quad &\mathrm{for} \quad 
	 \Delta_{\perp} \sim Q_s,
	 \\*[0.4cm]
	 {\displaystyle \frac{2 Q_A^4}{\Delta_{\perp}^2}
	\ln \frac{\Delta_{\perp}^2}{\Lambda^2}}
	 \quad &\mathrm{for} \quad 
	 \Delta_{\perp} \gg Q_s,
	\end{cases}
\end{align}
in which we clearly see the emergence of saturation at small momenta. More generally, $\mcal{S}_g(B)$ and hence $\mcal{G}^{(+)}(\Delta_{\perp})$ must be obtained at $\YP =\ln 1/\xP$ according to the discussion at the end of Sect.~\ref{sec:2jets}. In the general formula for $\xP$ in Eq.~\eqref{xp_and_beta} one must replace the diffractive mass $M_\chi^2$ with its two jets expression given in Eq.~\eqref{Mchi_2jets}, whereas one can neglect the momentum transfer $|t|$ since it is much smaller than $Q^2 \mkern-2mu+\mkern-2mu M_\chi^2$.

\section{2\,+\,1 jets: Soft gluon and incoherent gluon DTMD}
\label{sec:2_plus_g}

The $1/P_{\perp}^6$ spectrum obtained in Sect.~\ref{sec:2_hard_jets} indicates a rather strong fall-off as the hard momentum increases. However, this feature is true just for the leading order calculation in QCD perturbation theory. A much more efficient production mechanism  was worked out in detail in \cite{Iancu:2021rup,Iancu:2022lcw,Iancu:2023lel} for the case of coherent diffraction and the idea applies also to the problem under study. When the two hard fermion jets are accompanied by a third semi-hard gluon jet with transverse momentum $\bk_3$ such that $k_{3\perp} \sim Q_s \ll P_{\perp}$, what matters is not the small size $r$ of the $q\bar{q}$ pair but the large size $R \sim 1/k_{3\perp} \sim 1/Q_s$ of the whole $q\bar{q}g$ fluctuation. The scattering of such a configuration off the nucleus target is strong, so that the resulting hard dijet spectrum is not suppressed and falls only like $1/P_{\perp}^4$. Despite the fact that such a NLO contribution comes with an extra factor $\alpha_s(P_{\perp}^2)$, clearly it becomes the dominant one when $P_{\perp} \gg Q_s$, since the coupling decreases only logarithmically with increasing $P_{\perp}$. It is also important to realize that the gluon's longitudinal momentum must be rather soft. Indeed, its formation time $2 k_3^+/k_{3 \perp}^2$ must be of the order of the $\gamma^*$ lifetime $2 q^+/Q^2$, which in turn means that the splitting fraction $\vartheta_3 \equiv k_3^+/q^+$ must satisfy $\vartheta_3 \sim k_{3\perp}^2/Q^2 \sim Q_s^2/P_{\perp}^2 \ll 1$. Gluons with even smaller fractions are not relevant to our purposes, because they diminish the magnitude of the rapidity gap. We will refer to the third jet as either the semi-hard jet or the soft jet.

\begin{figure}
	\begin{center}
		\includegraphics[width=0.45\textwidth]{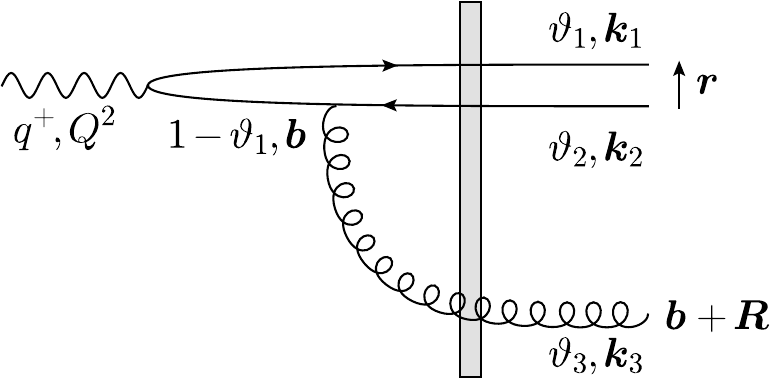}
		\hspace*{0.08\textwidth}
		\includegraphics[width=0.45\textwidth]{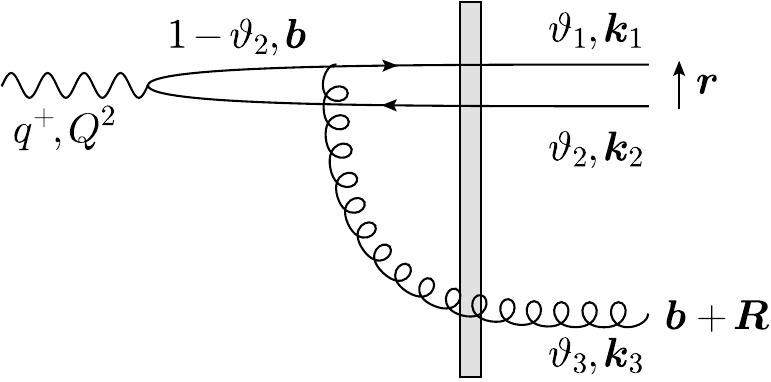}
	\end{center}
	\caption{\small Production of a hard quark-antiquark pair plus a soft gluon in DIS off a nucleus in the dipole picture. The scattering off the shockwave is taken only after the gluon radiation, since the large pair is formed after the gluon vertex.}
\label{fig:softg}
\end{figure}

\begin{figure}
	\begin{center}
	\includegraphics[width=0.65\textwidth]{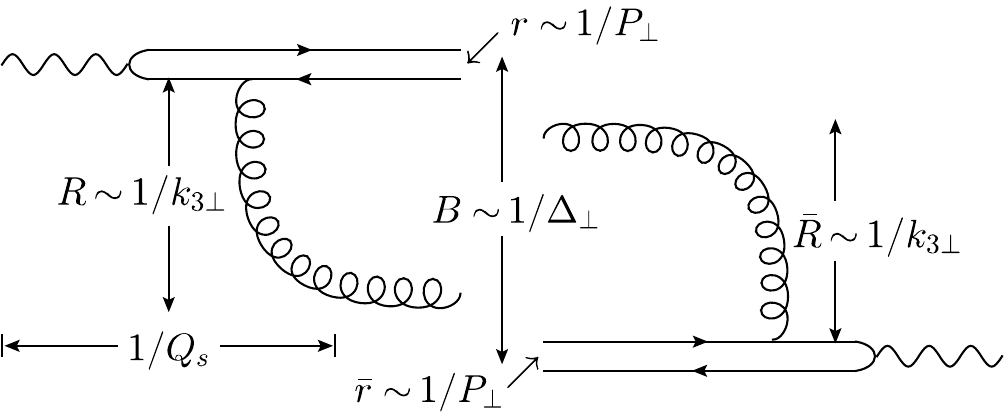}
	\end{center}
	\caption{\small A representative configuration which contributes to the incoherent diffractive production of a hard quark-antiquark pair plus a semi-hard gluon. The kinematic regime of interest is $P_{\perp} \gg \Delta_{\perp}, k_{3\perp}, Q_s$ which translates to the relation $r,\bar{r} \ll B, R,\bar{R}, 1/Q_s$ for the various sizes and distances in coordinate space. In the typical case $\Delta_{\perp} \sim k_{3\perp} \sim Q_s$, all the distances $B$, $R$ and $\bar{R}$ are of the order of $1/Q_s$.}
\label{fig:sizes_qqg}
\end{figure}

The full derivation for coherent diffraction has been given in \cite{Iancu:2022lcw} and here we will just present the results properly modified to describe an incoherent reaction. Let's start by being more precise on the relevant kinematics and the scattering. As we show in the amplitudes in Fig.~\ref{fig:softg}, after the decay of the virtual photon to the $q\bar{q}$ pair with size $r$, any of the two fermions can emit the soft gluon at a distance $R \gg r$ and thus the triparton fluctuation is effectively a gluon-gluon dipole with size $R$. It scatters eikonally off the nucleus target from which it gets a transverse momentum $\bm{\Delta}$ and a minus longitudinal momentum $\xP P_N^-$, with $\xP$ to be shortly given. The quark and antiquark carry corresponding fractions $\vartheta_1$ and $\vartheta_2$ of the plus longitudinal momentum $q^+$ of the virtual photon, so that $\vartheta_1+\vartheta_2 \simeq 1$, while $\vartheta_3 \ll 1$. The final transverse momenta $\bk_1$ of the quark and $\bk_2$ of the antiquark are hard and their imbalance receives contributions from both the gluon recoil and the momentum transfer, since transverse momentum requires $\bk_1+\bk_2 = -\bk_3 + \bm{\Delta}$. We stress that the typical scale for both $k_{3\perp}$ and $\Delta_{\perp}$ is $Q_s$, since it is set by the scattering. This is one of the features that distinguish the current 2\,+\,1 jets process not only from the corresponding coherent one, but also from the incoherent 2 jets process reviewed in Sections \ref{sec:2jets} and \ref{sec:2_hard_jets}. In both such problems there is only a single source of imbalance, the recoil gluon momentum in the first and the momentum transfer in the second. We find convenient to define $\bk_1= \bP + \vartheta_1(\bm{\Delta} - \bk_3)$ and  $\bk_2= -\bP + \vartheta_2(\bm{\Delta} - \bk_3)$, thus we shall use the hard momentum $\bP$ and the semi-hard momenta $\bm{\bk}_3$ and $\bm{\Delta}$ as the independent variables. In Fig.~\ref{fig:sizes_qqg} we give an illustration of a typical configuration of interest in coordinate space.

The hard factor is the same as the one for 2\,+\,1 jets in coherent diffraction, which further coincides with the one in inclusive dijet production. For definiteness, we consider the transverse sector for which we have  
\begin{align}
	\label{HTgij}
	H_T^{ij(g)} (\vartheta_1,\vartheta_2, \bP, \bar{Q})=
	\big(\vartheta_1^2 + \vartheta_2^2\big)
	\int	
	\frac{\rmd^2\bm{r}}{2\pi}\,
	\frac{\rmd^2\bar{\bm{r}}}{2\pi}\,
	e^{- i \bm{P} \cdot (\bm{r}-\bar{\bm{r}})}\,
	r^i \bar{r}^j\,
	\frac{\bm{r} \!\cdot\! \bar{\bm{r}}}{r \bar{r}}\,
	\bar{Q}\,
	K_{1}\big(\bar{Q}r\big)
	\bar{Q}\,
	K_{1}\big(\bar{Q}{\bar{r}}\big),
\end{align}
where the superscript $(g)$ denotes that the gluon is the soft parton in the configuration. The additional factor $r^i$ in the DA, when compared to Eq.~\eqref{sigmab}, is due to the vertex for the emission of a large-size gluon with polarization $i$ from the $q\bar{q}$ dipole with size $r$ (and similarly for the factor $\bar{r}^j$ in the CCA). The integrals in Eq.~\eqref{HTgij} are easily calculated and the hard factor can be written as the sum of a diagonal and a traceless part, more precisely 
\begin{align}
	\label{HTgij_dec}
	H_T^{ij(g)} (\vartheta_1,\vartheta_2,\bP, \bar{Q})=\,&\,
	\big(\vartheta_1^2 + \vartheta_2^2\big)
	\left[\frac{(P_{\perp}^4 +\bar{Q}^4)}{(P_{\perp}^2 +\bar{Q}^2)^4}
	\,\delta^{ij}
	- 
	\frac{4 \bar{Q}^2 P_{\perp}^2}{(P_{\perp}^2 +\bar{Q}^2)^4}\,
	\hat{P}^{ij}\right] 
	\nonumber
	\\*[0.1cm]
	\equiv \,&\, 
	H_T^{(g)}(P_{\perp},\bar{Q})
	\,\delta^{ij}
	+ H_{T}^{\prime(g)}(P_{\perp},\bar{Q}) 
	\,\hat{P}^{ij},
\end{align}
where for an arbitrary 2-dimensional vector $V^i$ we have defined the traceless symmetric tensor
\begin{align}
	\label{Vij}
	\hat{V}^{ij} = \frac{V^i V^j}{V_{\perp}^2} - \frac{\delta^{ij}}{2}.
\end{align}

The invariant mass of the combined system of the three outgoing partons (which are individually on-shell) for $P_{\perp} \gg k_{3\perp},\Delta_{\perp}$ and $\vartheta_3 \ll 1$ is rather simple and reads
\begin{align}
	\label{Mchi_3jets}
	M_{\chi}^2 = (k_1 + k_2 +k_3)^2 
	\simeq \frac{P_{\perp}^2}{\vartheta_1\vartheta_2}+ \frac{k_{3\perp}^2}{\vartheta_3}
	\qquad 
	\mathrm{for \mkern6mu 2\,+\,1 \mkern6mu jets}.
\end{align}
The splitting fraction $\xP$ and the diffractive variable $\beta$ can be specified by inserting Eq.~\eqref{Mchi_3jets} into the expressions in Eq.~\eqref{xbj_xp_beta}, where we further recall that $|t|$ can be omitted since it is much smaller than $Q^2 \mkern-1mu + \mkern-1mu M_{\chi}^2$ like in the 2 jets process. We further define $x$ as the fraction of the minus longitudinal momentum of the Pomeron that is transferred to the $q\bar{q}$ pair only, namely
\begin{align}
	\label{x_frac}
	x= 
	\frac{\bar{Q}^2+ P_{\perp}^2}{\bar{Q}^2 + P_{\perp}^2 + 
	(\vartheta_1 \vartheta_2/\vartheta_3) k_{3\perp}^2}=
	\frac{\bar{Q}^2+ P_{\perp}^2}{\bar{Q}^2}\,\beta
\end{align}
and we shall use the first equation in the above to make a change of variable from $\vartheta_3$ to $x$. By definition, $x$ regulates the rapidity separation between the hard dijet and the edge of the gap on the projectile side. When $\vartheta_3$ is of the order of $k_{3\perp}^2/P_{\perp}^2$, Eq.~\eqref{x_frac} implies that $x$ is of the order of, say, one-half and therefore the rapidity gap in the 2\,+\,1 jets process is only roughly a unit smaller than the one in the 2 jets one, assuming equal $\xbj$ and hard jets kinematics.

Now that we have introduced the appropriate variables, we can present the semi-hard factor $\mcal{G}^{ij}_{\rm inc}$. Despite the fact that it depends on the two vectors $\bk_3$ and $\bm{\Delta}$, we show in Appendix \ref{app:Gij} as a straightforward exercise that it has only a diagonal component, i.e.
\begin{align}
	\label{Gij_G}
	\mcal{G}^{ij}_{\rm inc}(x, x_{\mathbb{P}}, \bk_3, \bm{\Delta}) = 
	\frac{\delta^{ij}}{4}\,	
	\mcal{G}_{\rm inc}(x, x_{\mathbb{P}}, \bk_3, \bm{\Delta}),
\end{align}
where $\mcal{G}_{\rm inc}$ is simply twice the trace of $\mcal{G}^{ij}_{\rm inc}$ in matrix language and reads
\begin{align}
	\label{Ginc}
	\mcal{G}_{\rm inc}(x, x_{\mathbb{P}}, \bk_3, \bm{\Delta})=
	\frac{1}{\pi}
	\int	 
	&\,\frac{\rmd^2\bm{B}}{2\pi}\,
	\frac{\rmd^2\bm{R}}{2\pi}\,
	\frac{\rmd^2\bar{\bm{R}}}{2\pi}\,
	e^{- i \bm{\Delta} \cdot \bm{B}  
	- i \bk_3 \cdot (\bm{R}-\bar{\bm{R}})}
	\nn
	&\,\times \hat{R}^{ik}
	\hat{\bar{R}}^{ki}
	\mcal{M}^2 K_2(\mcal{M}R)
	\mcal{M}^2 K_2(\mcal{M}\bar{R})
	\mcal{W}_g
	(\bm{R},\bar{\bm{R}},\bm{B}).
\end{align}
Notice that $\hat{R}^{ik}
	\hat{\bar{R}}^{ki} = (1/2)\cos 2 \phi_{R\bar{R}}$, with $\phi_{R\bar{R}}$ being the angle between $\bm{R}$ and $\bar{\bm{R}}$. Needless to say, in exact analogy with Eq.~\eqref{wfund}, $\mcal{W}_g
	(\bm{R},\bar{\bm{R}},\bm{B})$ is the connected piece of the CGC correlator $\big \langle S_g (\bm{R}+\bb,\bb)\mkern1.5mu$$S_g (\bar{\bb},\bar{\bm{R}}+\bar{\bb}) \big \rangle$ for a homogeneous nucleus target, with $S_g$ the $S$-matrix in the adjoint representation. Clearly $\bb$ and $\bb+\bm{R}$ are the positions of the two legs of the gluon-gluon dipole in the DA, the first being the small $q\bar{q}$ pair and the second the soft gluon, and similarly for the CCA. Finally, the scale $\mcal{M}^2$ in Eq.~\eqref{Ginc} reads
\begin{align}
	\label{M2_k3}
	 \mcal{M}^2 \equiv \frac{x}{1-x}\,k_{3\perp}^2.
\end{align}
$\mcal{G}_{\rm inc}$ in Eq.~\eqref{Ginc} is strikingly similar to $\mcal{Q}_{\rm inc}$ in Eq.~\eqref{Qinc}. Leaving aside the scattering part which was already discussed just above, $\mcal{G}_{\rm inc}$ contains two factors (one in the DA and one in the CCA) of the type $\hat{R}^{ik} \mcal{M}^2 K_2(\mcal{M}R)$ which have their origin in the soft gluon emission from the small $q\bar{q}$ dipole, whereas $\mcal{Q}_{\rm inc}$ involves two factors of the type $(r^i/r)\mcal{M}^2 K_1(\mcal{M}R)$ for the quark emission. Hence, in analogy to Eq.~\eqref{qDTMD}, and by the inclusion of the appropriate color factor $N_c^2-1$, we naturally define the incoherent gluon DTMD
\begin{align}
  \label{gDTMD}
  \frac{\rmd xG_{\mathbb{P}}^{\rm inc}
  (x, x_{\mathbb{P}}, \bk_3, \bm{\Delta})}
  {\rmd^2 \bk_3\mkern1mu \rmd^2\bm{\Delta}}
  \equiv
  \frac{S_{\perp} (N_c^2-1)}{4 \pi^3}\,	
	\frac{\mcal{G}_{\rm inc}(x, x_{\mathbb{P}}, \bk_3, \bm{\Delta})}
	{2\pi(1-x)}.
\end{align}
Clearly it would be redundant to repeat the full discussion after Eq.~\eqref{qDTMD} at the end of Sect.~\ref{sec:SIDDIS}. We stress that we were able to take a parton, this time a gluon, from the projectile LCWF and view it as a part of the Pomeron because it was soft. The Pomeron splits into a pair of gluons, one of which carries a splitting fraction $1-x$ and transverse momentum $\bk_3$ and appears in the final state. The other gluon carries a fraction $x$ and transverse momentum $-\bk_3 +\bm{\Delta}$ which are transferred to the small $q\bar{q}$ dipole and put it on-shell.

\begin{figure}
	\begin{center}
	\includegraphics[width=0.43\textwidth]{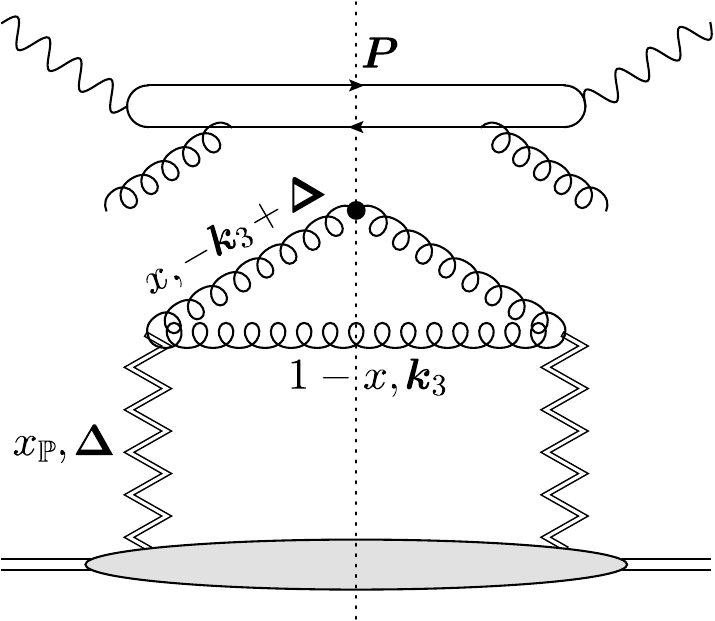}
	\end{center}
	\caption{\small TMD factorization in incoherent diffractive production of a hard quark-antiquark pair plus a soft gluon. The gluon emitted from the Pomeron is absorbed by the antiquark from the hard $q\bar{q}$ fluctuation of the virtual photon. (We show neither the similar diagram in which the gluon is absorbed by the quark nor the interference term.)}
\label{fig:TMD_gluon}
\end{figure}

The lack of a traceless part in the semi-hard factor in Eq.~\eqref{Gij_G}, means that only the diagonal piece of the hard factor in Eq.~\eqref{HTgij_dec} will contribute when we perform the contraction. That is, any potential angular dependence of the type $\bP \mkern-1mu\cdot \mkern-1mu \bm{\Delta}$  or $\bP \mkern-1mu\cdot \mkern-1mu \bk_3$ disappears\footnote{The dependence on the angle between $\bk_3$ and $\bm{\Delta}$ is not interesting and will be lost when integrating over $\bk_3$.} and therefore the only dependence on the hard momentum is via its magnitude $P_{\perp}$. Comparing with the coherent case \cite{Iancu:2022lcw} to easily infer the correct normalization factor, we finally arrive at the incoherent diffractive cross section for producing a hard quark-antiquark pair plus a soft gluon 
\begin{align}
	\label{cs_soft_g}
	\hspace{-0.1cm}
	\frac{\rmd\sigma^
	{\gamma_{\scriptscriptstyle \lambda}^* A\rightarrow q\bar q (g) X}}
	{\rmd \vartheta_1\mkern1mu
	\rmd \vartheta_2\mkern1mu
	\rmd^{2}\bP \mkern1mu
  	\rmd^{2}\bm{\Delta}\mkern1mu\mkern1mu
  	\rmd^2\bk_3\mkern1mu
  	\rmd \ln1/x} = 
  	\alpha_{\rm em} \alpha_s
  	\sum\mkern-1.5mu e_f^2\,
  	\delta_{\vartheta}\,
  	H_\lambda^{(g)} (\vartheta_1,\vartheta_2,P_{\perp}, \bar{Q})\,
  	\frac{\rmd xG_{\mathbb{P}}^{\rm inc}
  (x, x_{\mathbb{P}}, \bk_3, \bm{\Delta})}
  {\rmd^2 \bk_3\mkern1mu \rmd^2\bm{\Delta}},
\end{align} 
where $\delta_{\vartheta} \equiv \delta(1- \vartheta_1-\vartheta_2 -\vartheta_3) \simeq \delta(1- \vartheta_1-\vartheta_2)$ and with the incoherent gluon DTMD determined by Eqs.~\eqref{Ginc} and \eqref{gDTMD}. We recall that the transverse hard factor is given in Eq.~\eqref{HTgij_dec}, while for the corresponding longitudinal one we find 
\begin{align}
	\label{HLg}
	H_L^{(g)} (\vartheta_1,\vartheta_2,P_{\perp}, \bar{Q})=
	8 \vartheta_1\vartheta_2\,
	\frac{\bar{Q}^2 P_{\perp}^2}
	{(P_{\perp}^2 +\bar{Q}^2)^4}.
\end{align}
For the typical hard dijet kinematics, where $\vartheta_1,\vartheta_2$ are of the order of one-half and $P_{\perp} \sim \bar{Q}$, both hard factors behave like $1/P_{\perp}^4$, as asserted in the beginning of this section. The splitting of the Pomeron as described below Eq.~\eqref{gDTMD} and the TMD factorization which is manifest in Eq.~\eqref{cs_soft_g} are both illustrated in Fig.~\ref{fig:TMD_gluon}.

Before closing this section, we would like to emphasize that, since the momenta $\Delta_{\perp}$ and $k_{3\perp}$ are semi-hard, the corresponding sizes $B$ and $R,\bar{R}$ can be of the order of $1/Q_s$. Thus, in general, the CGC correlator $\mcal{W}_g
	(\bm{R},\bar{\bm{R}},\bm{B})$ must be computed without performing any expansion in order to keep track of all possible unitarity corrections\footnote{The same is true for $\mcal{W}_q
	(\bm{R},\bar{\bm{R}},\bm{B})$ which appears in the forthcoming section.}.

\section{2\,+\,1 jets: Soft quark and incoherent quark DTMD (\lowercase{again})}
\label{sec:2_plus_q}

Equally well to the case studied in the previous section, it is possible that the outgoing state is composed of a hard antiquark-gluon dijet with $k_{2\perp},k_{3\perp} \sim P_{\perp}$ and a semi-hard quark jet with $k_{1\perp} \sim Q_s \ll P_{\perp}$ (or, similarly, by a hard quark-gluon dijet and a semi-hard antiquark) \cite{Hauksson:2024bvv}. This is again a $q\bar{q}g$ configuration with large size $R \sim 1/k_{1\perp} \sim 1/Q_s$ and as such it scatters strongly with the target, eventually leading to a dijet spectrum which falls only like to $1/P_{\perp}^4$. It goes without saying that the quark must carry a splitting fraction $\vartheta_1 \sim k_{1\perp}^2/P_{\perp}^2 \ll 1$ of the longitudinal momentum $q^+$.

\begin{figure}
	\begin{center}
		\includegraphics[width=0.45\textwidth]{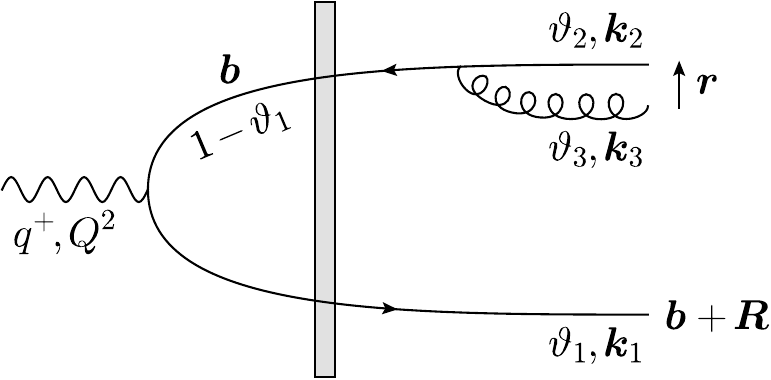}
		\hspace*{0.08\textwidth}
		\includegraphics[width=0.45\textwidth]{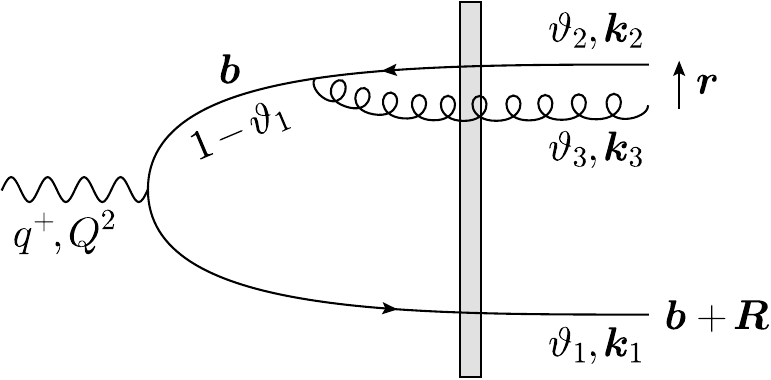}
	\end{center}
	\caption{\small Production of a hard antiquark-gluon pair plus a soft quark in DIS off a nucleus in the dipole picture via the channel $\bar{q}\to \bar{q} g$. The large pair is formed already at the photon vertex, thus the gluon may be radiated after the scattering off the shockwave (left panel) or before the scattering off the shockwave (right panel).}
\label{fig:fromqbar}
\end{figure}

\begin{figure}
	\begin{center}
		\includegraphics[width=0.45\textwidth]{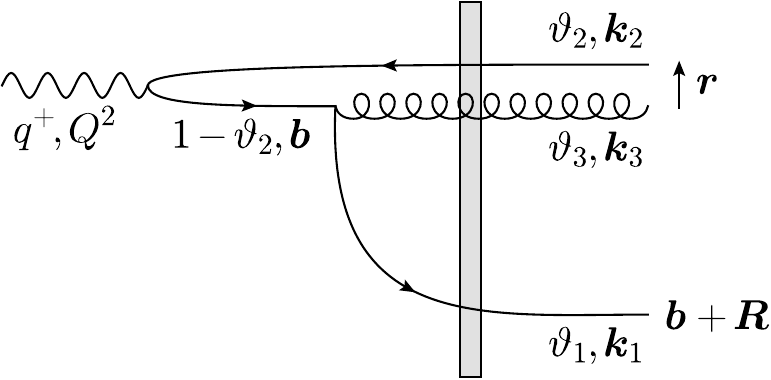}
	\end{center}
	\caption{\small Production of a hard antiquark-gluon pair plus a soft quark in DIS off a nucleus in the dipole picture via the channel $q\to q g$. The scattering off the shockwave is taken only after the gluon radiation, since the large pair is formed after the gluon vertex.}
\label{fig:fromq}
\end{figure}

The gluon, which now is hard, can be emitted either by the quark or by the antiquark. Since the emitters occur in the final state with very disparate momenta (the antiquark is hard, whereas the quark is semi-hard), the respective diagrams look different and they must be calculated separately. There are two scenarios which lead to the desired dynamics and final state. In the first, the virtual photon decays directly into a large $q\bar{q}$ pair of size $R$ and subsequently the semi-hard antiquark gives rise to a hard $\bar{q}g$ pair with size $r \ll R$. The scattering with the target nucleus can occur either before or after the gluon emission, since both the intermediate and final state configurations have a large size. The two corresponding amplitudes are shown in Fig.~\ref{fig:fromqbar}. The second scenario looks more similar to the soft gluon case analyzed in the previous section: the virtual photon decays into a hard $q\bar{q}$ pair of size $r$ and afterwards the (hard) quark gives rise to a soft quark at distance $R \gg r$ and a hard gluon. The latter together with the hard antiquark form a hard dijet of size $r$. Evidently, a favorable situation occurs only when the scattering takes place after the gluon emission and the amplitude is shown in Fig.~\ref{fig:fromq}. In both scenarios we always have to consider the scattering of a large $q\bar{q}$ dipole with its legs at $\bb+\bm{R}$ and $\bb$, while the kinematics of the final hard jets are $\bk_2 = \bP +\vartheta_2(\bm{\Delta}-\bk_1)$,  $\bk_3 = -\bP +\vartheta_3(\bm{\Delta} -\bk_1)$, $\vartheta_2+\vartheta_3 \simeq 1$, with $\Delta_{\perp},k_{1\perp},Q_s \ll P_{\perp}$. Finally, it is clear that the calculation of the cross section requires to also take into account the interference term between the two processes. A picture similar to the one given in Fig.~\ref{fig:sizes_qqg} holds in coordinate space, except that now it is the quark which plays the role of the semi-hard jet. 

The derivation follows closely the corresponding one for coherent diffraction \cite{Hauksson:2024bvv} and there is no reason to iterate all the details here, rather we shall directly present the final results. The hard factor remains identical and in the transverse sector we have 
\begin{align}
	\label{HTQij}
	H_T^{ij(q)}(\vartheta_2,\vartheta_3, P_{\perp},\tilde{Q}) = 
	\vartheta_2\, 
	\frac{(P_{\perp}^2 + \tilde{Q}^2)^2 + 
	\vartheta_2^2 \tilde{Q}^4+
	\vartheta_3^2 P_{\perp}^4}
	{P_{\perp}^2 (P_{\perp}^2 + \tilde{Q}^2)^3}\,
	\delta^{ij}
	\equiv
	H_T^{(q)}(\vartheta_2,\vartheta_3, P_{\perp},\tilde{Q})\,
	\delta^{ij},
\end{align}
with $\tilde{Q}^2 \equiv \vartheta_2\vartheta_3 Q^2$. Since the above tensor is diagonal\footnote{This property is true for each of the three individual terms. These arise from the square of the term with gluon emission from the antiquark, the square of the term with gluon emission from the quark and the respective interference term.}, we find again that the only dependence on the hard momentum is via its magnitude $P_{\perp}$. Now we move to the semi-hard factor which can be written in the form
\begin{align}
	\label{Qij_Q}
	\mcal{Q}^{ij}_{\rm inc}(x, x_{\mathbb{P}}, \bk_1, \bm{\Delta}) = 
	\frac{\delta^{ij}}{2}\,	
	\mcal{Q}_{\rm inc}(x, x_{\mathbb{P}}, \bk_1, \bm{\Delta})
	+ \hat{\mcal{Q}}^{ij}_{\rm inc}(x, x_{\mathbb{P}}, \bk_1, \bm{\Delta}),
\end{align}
where $\mcal{Q}_{\rm inc}$ is the quantity we already came across in Eq.~\eqref{Qinc}. We rewrite it here for our convenience in terms of the appropriate free and dummy variables, i.e.
\begin{align}
 	\label{QincR}
 	\mcal{Q}_{\rm inc}(x, x_{\mathbb{P}}, \bk_1, \bm{\Delta}) 
 	=
 	\frac{1}{2\pi}
 	\int 
 	&\,\frac{\rmd^2 \bm{B}}{2\pi}\,
	\frac{\rmd^2\bm{R}}{2\pi}\,
	\frac{\rmd^2\bar{\bm{R}}}{2\pi}
	e^{- i \bm{\Delta} \cdot \bm{B}
	- i \bk_1 \cdot (\bm{R}-\bar{\bm{R}})}\,
	\nn
	&\,\times\frac{\bm{R} \!\cdot\! \bar{\bm{R}}}{R \bar{R}}\,
	\mcal{M}^2
	K_{1}(\mcal{M}R)
	\mcal{M}^2
	K_{1}(\mcal{M}{\bar{R}})
	\mcal{W}_q(\bm{R},\bar{\bm{R}},\bm{B}).
\end{align}
We recall that $\mcal{W}_q
	(\bm{R},\bar{\bm{R}},\bm{B})$ is the connected piece of the CGC correlator $\big \langle S (\bm{R}+\bb,\bb)\mkern1.5mu$$S (\bar{\bb},\bar{\bm{R}}+\bar{\bb}) \big \rangle$ for a homogeneous nucleus target, with $S$ the $S$-matrix in the fundamental representation, whereas now
\begin{align}
	\label{M2_k1}
	 \mcal{M}^2 \equiv \frac{x}{1-x}\,k_{1\perp}^2.
\end{align}
$\hat{\mcal{Q}}^{ij}_{\rm inc}$ in Eq.~\eqref{Qij_Q} is traceless and as such it does not contribute to the cross section since it must be contracted with the hard tensor in Eq.~\eqref{HTQij} which is diagonal. Rewriting Eq.~\eqref{qDTMD} as
\begin{align}
  \label{qDTMDk}
  \frac{\rmd xq_{\mathbb{P}}^{\rm inc}
  (x, x_{\mathbb{P}}, \bk_1, \bm{\Delta})}
  {\rmd^2 \bk_1 \rmd^2\bm{\Delta}}
  \equiv
  \frac{S_{\perp} N_c}{4 \pi^3}\,	
	\frac{\mcal{Q}_{\rm inc}(x, x_{\mathbb{P}}, \bk_1, \bm{\Delta})}
	{2\pi(1-x)}
\end{align}
and following \cite{Hauksson:2024bvv}, we obtain the incoherent diffractive cross section for producing a hard antiquark-gluon pair plus a soft quark 
\begin{align}
	\label{cs_soft_q}
	\hspace{-0.3cm}
	\frac{\rmd\sigma^
	{\gamma_{\scriptscriptstyle \lambda}^* A\rightarrow (q)\bar q g X}}
	{\rmd \vartheta_2\mkern1mu
	\rmd \vartheta_3\mkern1mu
	\rmd^{2}\bP \mkern1mu
  	\rmd^{2}\bm{\Delta}\mkern1mu\mkern1mu
  	\rmd^2\bk_1\mkern1mu
  	\rmd \ln1/x} = 
  	2\alpha_{\rm em} \alpha_s C_F
  	\sum\mkern-1.5mu e_f^2\,
  	\delta_{\vartheta}
  	H_\lambda^{(q)} (\vartheta_2,\vartheta_3,P_{\perp}, \tilde{Q})
  	\frac{\rmd xq_{\mathbb{P}}^{\rm inc}
  (x, x_{\mathbb{P}}, \bk_1, \bm{\Delta})}
  {\rmd^2 \bk_1\mkern1mu \rmd^2\bm{\Delta}},
\end{align} 
where $\delta_{\vartheta} \equiv \delta(1- \vartheta_1-\vartheta_2 -\vartheta_3) \simeq \delta(1- \vartheta_2-\vartheta_3)$. The incoherent quark DTMD is determined by Eqs.~\eqref{QincR} and \eqref{qDTMDk}, the transverse hard factor is given in Eq.~\eqref{HTQij}, while the longitudinal one reads 
\begin{align}
	\label{HLq}
	H_L^{(q)} (\vartheta_2,\vartheta_3,P_{\perp}, \tilde{Q})=
	4 \vartheta_2^2\mkern1mu \vartheta_3\,
	\frac{\tilde{Q}^2}
	{(P_{\perp}^2 +\tilde{Q}^2)^3}.
\end{align}	
Like in the soft gluon sector, the hard factors behave like $1/P_{\perp}^4$ for typical kinematics of the hard dijet.

\begin{figure}
	\begin{center}
	\includegraphics[width=0.43\textwidth]{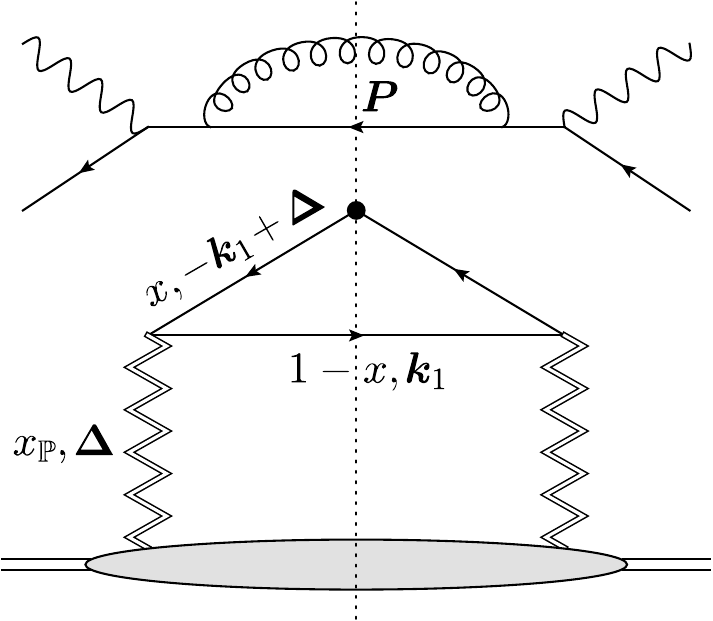}
	\hspace*{0.08\textwidth}
	\includegraphics[width=0.43\textwidth]{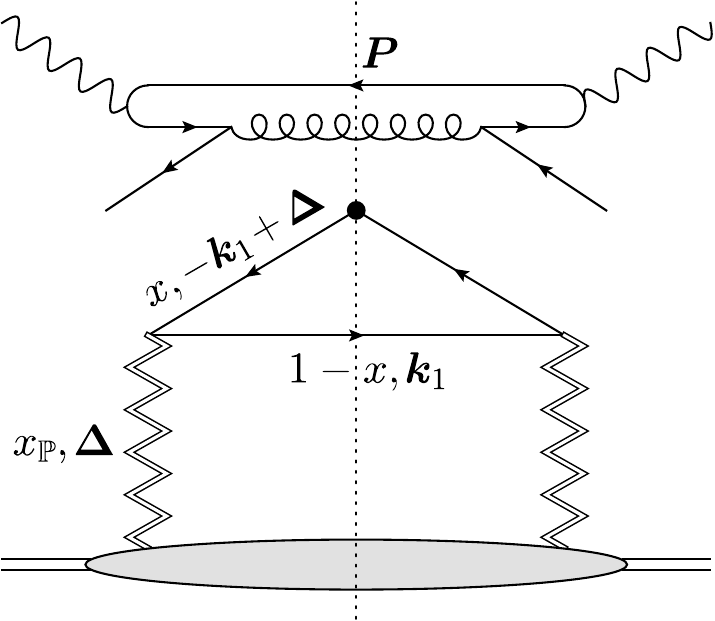}
	\end{center}
	\caption{\small TMD factorization in incoherent diffractive production of a hard antiquark-gluon pair plus a soft quark. Left panel: The antiquark emitted from the Pomeron absorbs the virtual photon and then gives rise to the hard pair. Right panel: The quark from the hard $q\bar{q}$ fluctuation of the virtual photon annihilates the antiquark emitted from the Pomeron to produce a hard gluon. The interference diagram is not shown.}
\label{fig:TMD_quark}
\end{figure}

We conclude this section by emphasizing that once more we have managed to take a parton, this time the quark, from the projectile LCWF and view it as a part of the Pomeron because it was soft. The Pomeron splits into a $q\bar{q}$ pair: the quark carries a splitting fraction $1-x$ and transverse momentum $\bk_1$ and appears in the final state, whereas the antiquark carries a fraction $x$ and transverse momentum $-\bk_1 +\bm{\Delta}$ which are transferred to the hard $\bar{q}g$ pair and put it on-shell. This picture and the TMD factorization in Eq.~\eqref{cs_soft_q} are illustrated in Fig.~\ref{fig:TMD_quark}.

\section{Study of the Incoherent DTMD\lowercase{s} and DPDF\lowercase{s}}
\label{sec:distributions}

One of our goals in this work is to make a direct comparison between the exclusive two jets cross section presented in Sect.~\ref{sec:2_hard_jets} and the 2\,+\,1 jets cross sections derived in Sections \ref{sec:2_plus_g} and \ref{sec:2_plus_q}. Before doing so, we must first integrate each of  the 2\,+\,1 jets cross sections over the kinematics of the soft jet. When we perform the integration over the semi-hard transverse momentum, we will naturally arrive to the incoherent gluon and quark DPDFs. The integration over the longitudinal momentum, will be done in a suitable way later on in Sect.~\ref{sec:mingap}. Hence it becomes necessary to understand how the gluon and quark DTMDs depend on the various variables, particularly on the transverse momentum $\bk$ of the semi-hard parton (denoted by $\bk_3$ for the gluon in Eq.~\eqref{gDTMD} and by $\bk_1$ for the quark in Eq.~\eqref{qDTMDk}) and on the splitting fraction $x$.

\subsection{The Incoherent DTMDs}
\label{sec:dtmds}

Obviously it suffices to study the functions $\mcal{G}_{\rm inc}(x, x_{\mathbb{P}}, \bk, \bm{\Delta})$ and $\mcal{Q}_{\rm inc}(x, x_{\mathbb{P}}, \bk, \bm{\Delta})$ given in Eqs.~\eqref{Ginc} and \eqref{QincR} respectively. We will assume that an averaging over the angle between the momenta $\bm{\Delta}$ and $\bk_3$ has been done, so that the dependence on these two vectors reduces to one on the corresponding magnitudes. We shall also take $\Delta_{\perp}$ to be fixed at a value of the order of $Q_s$, but we underline that a thorough analysis on the $\Delta_{\perp}$-dependence of the incoherent DPDFs will be presented in the next subsection. Finally, it is evident that $\xP$ enters as the scale at which the CGC correlators $\mcal{W}_g$ and $\mcal{W}_q$ should be evaluated and therefore the relative dependence comes predominantly through the saturation momentum $Q_s(\xP)$.

It is essential to realize that $Q_s(\xP)$ is not exactly the relevant saturation scale so long as the momentum $k_{\perp}$ is concerned. The exponentially falling Bessel functions $K_2(\mcal{M}R)$ and $K_1(\mcal{M}R)$ together with the oscillations due to the Fourier transform indicate that the effective maximum value of the size $R$ scales like $R_{\max}^2 \sim  (1-x)/k_{\perp}^2$ \cite{Iancu:2021rup,Iancu:2022lcw,Hauksson:2024bvv}. Physically this means that the formation of a soft parton with transverse momentum $k_{\perp}$ is not complete when $x$ is not very small and gets significantly suppressed when $1-x\ll 1$. The integration will be sensitive to unitarity corrections in the correlators $\mcal{W}_g$ and $\mcal{W}_q$ if and only if $R_{\max}^2 \sim 1/Q_s^2(\xP)$, that is, when $k_{\perp}^2 \sim \tilde{Q}_s^2(x,\xP)$ where we have defined the effective gluon saturation momentum \cite{Iancu:2021rup,Iancu:2022lcw,Hauksson:2024bvv}
\begin{align}
	\label{Qstilde}
	\tilde{Q}_s^2(x,\xP) \equiv (1-x) Q_s^2(\xP).
\end{align} 
An exactly analogous relation can be written for the quark saturation scales $\tilde{Q}_{s,q}^2(x,\xP)$ and $Q_{s,q}^2(x,\xP)$.

Even though we shall not perform any analytic calculations here, we can show by simple estimates that both $\mcal{G}_{\rm inc}$ and $\mcal{Q}_{\rm inc}$ exhibit a power law tail of the type $\tilde{Q}_s^4/k_{\perp}^4$ (in the MV model) for $k_\perp \gg \tilde{Q}_s$, whereas $\mcal{G}_{\rm inc}$ saturates and $\mcal{Q}_{\rm inc}$ vanishes for $k_\perp \ll \tilde{Q}_s$. We also find that $\mcal{Q}_{\rm inc}$ vanishes linearly with $x$ when $x\ll 1$, like in coherent diffraction \cite{Hauksson:2024bvv}. On the contrary, $\mcal{G}_{\rm inc}$ and $\mcal{Q}_{\rm inc}$ remain non-zero when $1-x \ll 1$, unlike in coherent diffraction\footnote{The respective quantities in the coherent case vanish like $(1-x)^2$ and $(1-x)$, cf.~Eqs.~(5.8) and (5.13) in \cite{Iancu:2022lcw} and (6.2) and (B.11) in \cite{Hauksson:2024bvv}.}. We shall give the technical explanation for the source of such a difference at the end of Appendix \ref{app:gDPDF}. The qualitative behavior of $\mcal{G}_{\rm inc}$ can be summarized in 
\begin{align}
	\label{G_and_Q_qual}
	\mcal{G}_{\rm inc}
	\sim
	\frac{1}{\pi N_c^2\mkern1.5mu Q_s^2} \times 
	\begin{cases}
	1 
	&\mathrm{for} \quad 
	k_{\perp} \ll \tilde{Q}_s(x,\xP),
	 \\*[0.4cm]
 	\displaystyle\frac{\tilde{Q}_s^4(x,\xP)}{k_{\perp}^4}
 	&\mathrm{for} \quad
 	k_{\perp} \gg \tilde{Q}_s(x,\xP),
	\end{cases}
\end{align}
where we have neglected possible logarithmic dependences and we have assumed that $\Delta_{\perp}$ is fixed, and similarly for $(1/x) \mcal{Q}_{\rm inc}$ with the difference that it vanishes at small $k_{\perp}$. Notice that $\mcal{G}_{\rm inc}$ and $\mcal{Q}_{\rm inc}$ have mass dimension -2 and thus when $\Delta_{\perp} \sim Q_s$ we expect them to scale with $1/Q_s^2$ as explicitly shown in the above equation.

\begin{figure}
	\begin{center}
	\includegraphics[width=0.48\textwidth]{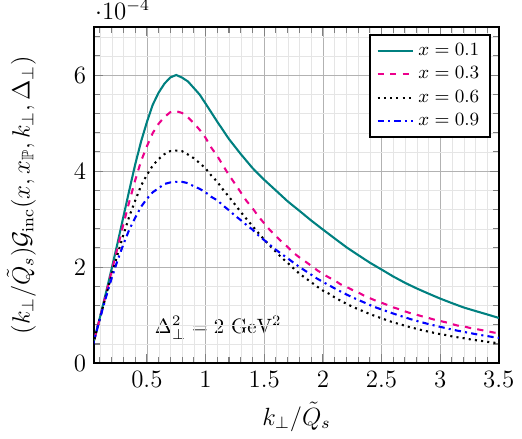}
	\hspace*{0.02\textwidth}
	\includegraphics[width=0.48\textwidth]{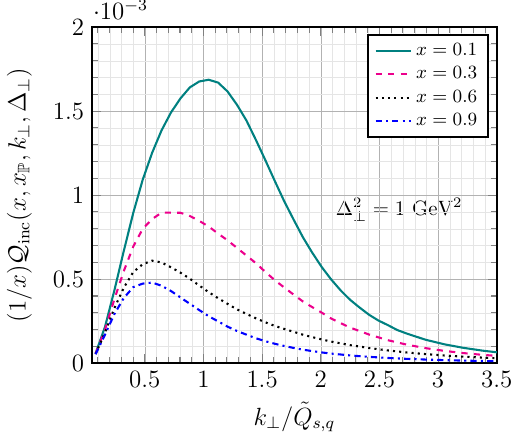}
	\end{center}
	\caption{\small Left panel: The function $\mcal{G}_{\rm inc}(x, x_{\mathbb{P}}, k_{\perp}, \Delta_{\perp})$ given in Eq.~\eqref{Ginc} and which appears in the definition of the incoherent gluon DTMD in Eq.~\eqref{gDTMD} multiplied with $k_{\perp}/\tilde{Q}_s(x)$ and plotted in terms of $k_{\perp}/\tilde{Q}_s(x)$ for fixed $\Delta_{\perp}$ and for various values of $x$. Right panel: The function $\mcal{Q}_{\rm inc}(x, x_{\mathbb{P}}, k_{\perp}, \Delta_{\perp})$ given in Eq.~\eqref{QincR} and which appears in the definition of the incoherent quark DTMD in Eq.~\eqref{qDTMDk} multiplied with the ``singular'' factor $1/x$ and plotted in terms of $k_{\perp}/\tilde{Q}_{s,q}(x)$ for fixed $\Delta_{\perp}$ and for various values of $x$. In both cases an average over the angle between $\bk$ and $\bm{\Delta}$ has been performed and the scattering amplitude is given by the MV model with a gluon saturation scale $Q_s^2 = 2\mkern1.5mu{\rm GeV}^2$.}
\label{fig:DTMDs}
\end{figure}

The above discussion suggests that both $k_{\perp} \mcal{G}_{\rm inc}$ 
and  $k_{\perp} \mcal{Q}_{\rm inc}$ should display a clear maximum at $k_{\perp} \sim \tilde{Q}_s$, followed by a rather steep tail at large momenta. This is indeed what we observe in the numerical results presented in Fig.~\ref{fig:DTMDs} for a fixed $\Delta_{\perp}$ of the order of the saturation momentum\footnote{In all the numerical calculations we define the saturation momentum according to $S_g(R = 2/Q_s)=1/2$. This is slightly different from the definition in Eqs.~\eqref{SgMV} and \eqref{QAQs} which is more convenient for the analytical calculations. We also regularize the infrared behavior of the MV model in Eq.~\eqref{SgMV} by letting $\ln(4/R^2 \Lambda^2) \to 2 \ln [(2/R\Lambda) + e].$}. Furthermore, we have numerically checked that such properties remain valid for different values of $\Delta_{\perp}$. The main change that occurs with increasing $\Delta_{\perp}$ is that the whole spectrum moves to larger momenta, albeit very slowly and such a feature is totally unimportant to our purposes. 

\comment{\caption{\small The quantities $\mcal{G}_{\rm inc}(x, x_{\mathbb{P}}, k_{\perp}, \Delta_{\perp})$ (left) and $\mcal{Q}_{\rm inc}(x, x_{\mathbb{P}}, k_{\perp}, \Delta_{\perp})$ (right) given in Eqs.~\eqref{Ginc} and \eqref{QincR} and which appear in the definition of the incoherent gluon and quark DTMDs in Eqs.~\eqref{gDTMD} and \eqref{qDTMDk} plotted as functions of $k_{\perp}/\tilde{Q}_s(x)$ for fixed $\Delta_{\perp}$ and for various values of $x$. An average over the angle between $\bk$ and $\bm{\Delta}$ has been performed. For the gluon we have multiplied  with $k_{\perp}/\tilde{Q}_s(x)$, whereas for the quark we have multiplied with the ``singular'' factor $1/x$, cf.~the discussion below Eq.~\eqref{Qstilde}. 
		The scattering amplitude is given by the MV model with a gluon saturation scale $Q_s^2 = 2\mkern1.5mu{\rm GeV}^2$.}}

\subsection{The Incoherent DPDFs}
\label{sec:dpdfs}

Not unexpectedly, we now define the incoherent DPDFs by integrating the corresponding DTMDs over the transverse momentum up to the scale of interest, which in our problem is the hard momentum $P_{\perp}$. Thus, for the incoherent gluon DPDF we have
\begin{align}
	\label{DPDF_def}
	\hspace{-0.7cm}
  \frac{\rmd xG_{\mathbb{P}}^{\rm inc}
  (x, x_{\mathbb{P}}, P_{\perp}, \Delta_{\perp})}
  {\rmd^2\bm{\Delta}}
  \equiv
  \int^{P_\perp}\!\!\! \dif^2 \bk\,\frac{\rmd xG_{\mathbb{P}}^{\rm inc}
  (x, x_{\mathbb{P}}, \bk, \bm{\Delta})}
  {\rmd^2 \bk \mkern1mu\rmd^2\bm{\Delta}}
  = \frac{S_{\perp} (N_c^2-1)}{4 \pi^3}\!
  \int^{P_\perp}\!\!\! \dif^2 \bk\,
  \frac{\mcal{G_{\rm inc}} (x, x_{\mathbb{P}}, \bk, \bm{\Delta})}{2\pi (1-x)}\end{align}
and similarly for the quark one, but with a color factor $N_c$. These are the analogues of the well-known DPDFs appearing in the coherent case \cite{Golec-Biernat:2001gyl,Hauksson:2024bvv}. 

Given the high momentum behavior of the DTMDs which is reflected in Eq.~\eqref{G_and_Q_qual}, we immediately see that if we integrate up to infinity in Eq.~\eqref{DPDF_def}, we must subtract a correction which falls like $1/P_{\perp}^2$, but such a rather weak dependence can be safely neglected\footnote{In fact, we have been systematically ignoring corrections of this order, since we have assumed throughout the whole calculation that $k_{\perp},Q_s, \Delta_{\perp} \ll P_{\perp}$.}. Logarithmic corrections, which are a bit more sizeable, can arise due to the DGLAP evolution from the semi-hard scale $Q_s$ up to the hard scale $P_{\perp}$ \cite{Iancu:2021rup,Iancu:2022lcw,Iancu:2023lel,Hauksson:2024bvv}. In principle they must be taken into account, but the observable of interest, to be defined in Sect.~\ref{sec:mingap}, is mostly sensitive to values of $x$ around one-half, a regime in which DGLAP is not expected to cause significant changes. Thus, even though this is not perfectly true, we will take the DPDFs to be $P_{\perp}$-independent in this work.

The $k_{\perp}$-integration is controlled by values in the regime $k_{\perp} \sim \tilde{Q}_s$ and simply brings in a factor $\tilde{Q}_s^2 = (1-x) Q_s^2$, which cancels the $1/Q_s^2$ prefactor in Eq.~\eqref{G_and_Q_qual} and the $1/(1-x)$ factor in the DTMD definitions in Eqs.~\eqref{gDTMD} and \eqref{qDTMDk}. Without doubt, the incoherent quark DPDF will also carry a factor proportional to $x$ inherited from the respective DTMD.

\begin{figure}
	\begin{center}
	\includegraphics[width=0.48\textwidth]{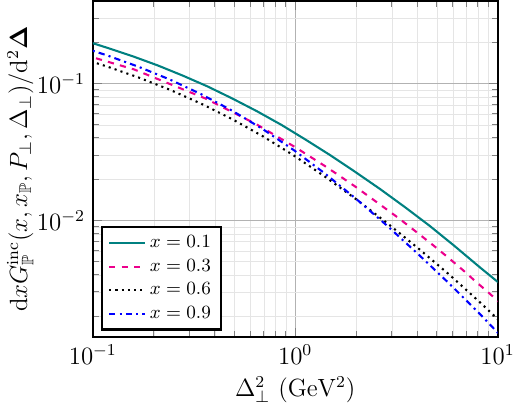}
	\hspace*{0.02\textwidth}
	\includegraphics[width=0.48\textwidth]{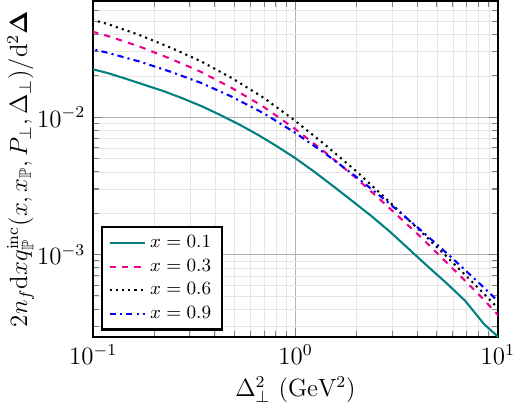}
	\end{center}
	\caption{\small The incoherent gluon (left) and quark (right) DPDFs (the latter multiplied by 2$n_f$, where $n_f=3$ is the number of flavors) as functions of the momentum transfer $\Delta_{\perp}^2$ for various values of $x$. The hard scale $P_{\perp}^2$, cf.~Eq.~\eqref{DPDF_def}, has been taken equal to infinity, the scattering amplitude is given by the MV model with a gluon saturation scale $Q_s^2 = 2\mkern1.5mu{\rm GeV}^2$ and a factor $S_{\perp}/4\pi^3$ has been neglected.}
\label{fig:DPDFs}
\end{figure} 

Finally, employing the Gaussian approximation (in the large-$N_c$ limit) in order to express $\mcal{W}_g$ in terms of $\mcal{S}_g$ and assuming the MV model for the latter, we work out the $\Delta_{\perp}$-dependence of the gluon DPDF  in full detail in Appendix \ref{app:gDPDF} separately in the regimes $\Delta_{\perp} \ll Q_s$ and $\Delta_{\perp} \gg Q_s$ and close to the end-points $x=0$ and $x=1$, to arrive at Eqs.~\eqref{appeq:DPDF_0low}, \eqref{appeq:DPDF_0high}, \eqref{appeq:DPDF_1low} and \eqref{appeq:DPDF_1high}. Not surprisingly, it saturates (logarithmically) at small momenta, while it shows the, familiar by now, $Q_s^4/\Delta_{\perp}^4$ tail at high momenta.

\begin{figure}
	\begin{center}
	\includegraphics[width=0.48\textwidth]{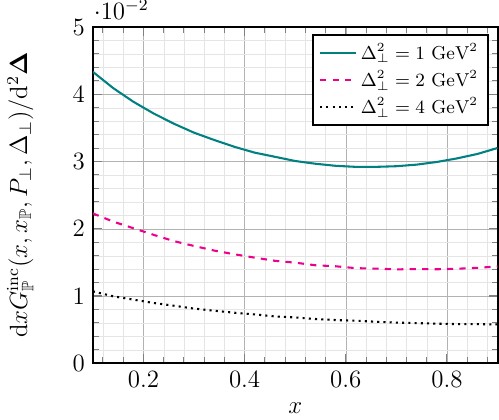}
	\hspace*{0.02\textwidth}
	\includegraphics[width=0.48\textwidth]{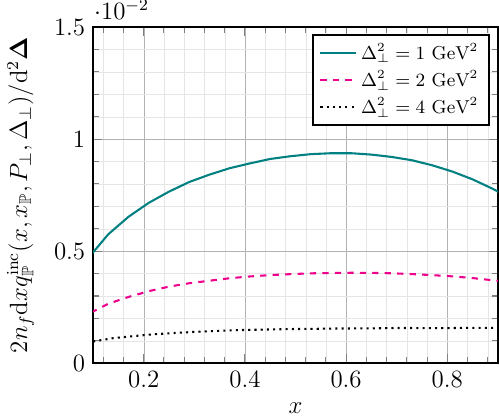}
	\end{center}
	\caption{\small The incoherent gluon (left) and quark (right) DPDFs (the latter multiplied by 2$n_f$) as functions of the splitting fraction $x$ for various values of $\Delta_{\perp}^2$. The hard scale $P_{\perp}^2$, cf.~Eq.~\eqref{DPDF_def}, has been taken equal to infinity, the scattering amplitude is given by the MV model with a gluon saturation scale $Q_s^2 = 2\mkern1.5mu{\rm GeV}^2$ and a factor $S_{\perp}/4\pi^3$ has been neglected.}
\label{fig:DPDFs_x}
\end{figure}

Neglecting some constant factors and logarithms, we can summarize the qualitative behavior of the two incoherent DPDFs in the piecewise expression
\begin{align}
	\hspace{-0.6cm}
	\label{DPDFs_qual}
	\frac{\rmd xG_{\mathbb{P}}^{\rm inc}
  (x, x_{\mathbb{P}}, P_{\perp}, \Delta_{\perp})}
  {\rmd^2\bm{\Delta}},
  \frac{N_c}{x}\,\frac{\rmd xq_{\mathbb{P}}^{\rm inc}
  (x, x_{\mathbb{P}}, P_{\perp}, \Delta_{\perp})}
  {\rmd^2\bm{\Delta}}
	\sim
	\frac{S_{\perp}}{8\pi^4}
	\times
	\begin{cases}
	\displaystyle \ln \frac{Q_s^2(\xP)}{\Lambda^2}
	&\mathrm{for} \quad 
	\Delta_{\perp} \ll Q_s(\xP),
	 \\*[0.4cm]
 	\displaystyle\frac{Q_s^4(\xP)}{\Delta_{\perp}^4}
 	&\mathrm{for} \quad
 	\Delta_{\perp} \gg Q_s(\xP).
	\end{cases}
\end{align}
We point out that $Q_s(\xP)$ is the relevant saturation scale so long as the momentum $\Delta_{\perp}$ is concerned. In Figs.~\ref{fig:DPDFs} and \ref{fig:DPDFs_x} we show our numerical results, which confirm all the aforementioned features regarding the dependence of the DPDFs on both $\Delta_{\perp}$ and $x$. We also stress that both DPDFs keep increasing logarithmically as we decrease $\Delta_{\perp}$ below $Q_s$.  

\section{Diffractive Dijet production with a minimum rapidity gap}
\label{sec:mingap}

The integration over the transverse momentum of the semi-hard jet in the 2\,+\,1 jet cross sections in Eqs.~\eqref{cs_soft_g} and \eqref{cs_soft_q} is trivial: we just need to replace the incoherent DTMDs with the corresponding DPDFs. Still, we cannot carry out yet a direct comparison with the 2 jets cross section in Eq.~\eqref{sigmaave} because the phase-spaces are different. While all cross sections depend on $P_{\perp}$, $\Delta_{\perp}$ and the hard jet splitting fractions (notice that only one of the two is independent, since their sum is equal or approximately equal to one), the 2\,+\,1 jets cross sections also depend on the minus longitudinal momentum splitting fraction $x$, whose definition involves the kinematics of the semi-hard jet. A suitable way to proceed is to impose a minimum value for the rapididy gap $\YP$ of the diffractive process, independently of the number of jets appearing in the outgoing state. We shall shortly see that this requires to integrate the 2\,+\,1 jets cross section over $x$ in a certain kinematical regime.

We shall take $s$ and $Q^2$ to be fixed, so that the total available rapidity phase space $Y_{\scriptscriptstyle \rm Bj} = \ln 1/\xbj$ is fixed too. Then Eq.~\eqref{xbj_xp_beta} implies that imposing a minimum value for $\YP$ (or a maximum for $\xP$) is equivalent to imposing a maximum value for $Y_{\beta}$ (or a minimum for $\beta$), the rapidity interval occupied by the projectile diffractive system. To be more precise, let's take a center of mass energy (squared) $s= 8800\,{\rm GeV}^2$, hopefully achievable at the EIC in DIS off heavy nuclei, and a photon virtuality $Q^2=20\,{\rm GeV}^2$ which lead to $\xbj \simeq 2.3 \times 10^{-3}$, that is $Y_{\scriptscriptstyle \rm Bj} \simeq 6.1$. We shall consider two cases for the minimum rapidity gap according to
\begin{align}
	\label{mingap1} 
	\YP^{\min} \simeq 4.5
	\,\, \Leftrightarrow \,\,
	\xP^{\max} \simeq 1.1 \times10^{-2} 
	\quad | \quad
	Y_{\beta}^{\max} \simeq 1.6
	\,\, \Leftrightarrow \,\,
	\beta_{\min} = 0.2 
\end{align} 
and
\begin{align}
	\label{mingap2} 
	\YP^{\min} \simeq 4
	\,\, \Leftrightarrow \,\,
	\xP^{\max} \simeq 1.8\times10^{-2} 
	\quad | \quad
	Y_{\beta}^{\max} \simeq 2.1
	\,\, \Leftrightarrow \,\,
	\beta_{\min} = 0.12. 
\end{align} 
The CGC correlators are to be computed at the longitudinal scale $\xP$ and the above values indicate that saturation effects should be relevant.

 Finally, to be definite, but without any loss of generality, we will assume that $\vartheta_{h1}=\vartheta_{h2}=1/2$, where $\vartheta_{h1}$ and $\vartheta_{h2}$ are the splitting fractions of the two forward hard jets. Thus, it remains to study the cross sections in terms of the hard jet momentum $P_{\perp}^2$ and the semi-hard momentum transfer $\Delta_{\perp}^2$. Before doing that, we first need to see what the minimum gap constraint means for the 2 jets and the 2\,+\,1 jets configurations separately.

\paragraph{2 jets:} Intuitively we understand that as $P_{\perp}^2$ increases, the diffractive mass of the projectile system increases and the gap decreases. Hence, since all of the remaining kinematics is already fixed, Eq.~\eqref{beta_2jets} means that the constraint for a minimum rapidity gap translates to a maximum allowed value for the hard transverse momentum, more precisely
\begin{align}
	\label{Pperp_max_beta}
	 P_{\perp\max}^2 =
	 \frac{1-\beta_{\min}}{\beta_{\min}}\,
	 \bar{Q}^2.
\end{align}
With $\vartheta_1 = \vartheta_2 =1/2$ and $Q^2=20\,{\rm GeV}^2$ one readily finds
\begin{align}
	\label{Pperp_max_numbers}
	 P_{\perp\max}^2  = 
	 \begin{cases}
	 	20\,{\rm GeV}^2 \quad &\mathrm{for} \quad \beta_{\min} = 0.2\phantom{1}
	 	\quad (\YP=4.5)
	 	\\*[0.2cm]
	 	36.7\,{\rm GeV}^2 \quad &\mathrm{for} \quad \beta_{\min} = 0.12
	 	\quad (\YP=4).
	 \end{cases}
\end{align}
We make it clear that, both the above discussion for 2 jets and the following one for 2\,+\,1 jets do not depend on the particular choice for $\Delta_{\perp}^2$, so long as the latter satisfies our assumption $\Delta_{\perp}^2 \ll P_{\perp}^2$.\\

\paragraph{2\,+\,1 jets:} Now for a given $P_{\perp}^2$ the gap is not fixed
due to the presence of the third, soft, jet, whose momentum can vary. Recalling Eq.~\eqref{x_frac}, let us rewrite the diffractive variable $\beta$ for 2\,+\,1 jets in the form
\begin{align}
	\label{beta_2plus1}
	\beta = \frac{Q^2}{Q^2 + P_{\perp}^2/\vartheta_{h1} 
	\vartheta_{h2} + 
	k_{\perp}^2/\vartheta},
\end{align} 
where $k_{\perp}^2$ and $\vartheta$ pertain to the transverse momentum and the splitting fraction of the semi-hard jet. It goes without saying that for the same hard jet kinematics the gap is smaller than the one with only 2 jets. Imposing a minimum value for $\beta$ leads to a condition with a clear physical meaning: $k_{\perp}^2/\vartheta$ does not get too large, because otherwise the semi-hard jet will start destroying the rapidity gap. 

Since it has been convenient to write the cross sections as a function of $x$, we also see that a minimum value of $\beta$ results in a minimum value for $x$. Indeed, using the second equality in Eq.~\eqref{x_frac} we immediately find that $x_{\rm min}$ is an increasing function of $P_{\perp}^2$, namely
\begin{align}
	\label{xmin}
	x_{\rm min} = \frac{Q^2 + P_{\perp}^2/\vartheta_{h1} 
	\vartheta_{h2}}{Q^2}\,\beta_{\rm min}.
\end{align}
On the other hand, there is no additional constraint for $x$ from above, i.e.~it can take values close to one\footnote{The derivation of the 2\,+\,1 jets cross sections in the regime of interest allows $\vartheta$ to be parametrically larger than $k_{\perp}^2/P_{\perp}^2$ (but not as large as $k_{\perp}/P_{\perp}$) \cite{Iancu:2022lcw,Hauksson:2024bvv} and then \eqref{x_frac} indicates that $x$ can approach one from below.}. It is also instructive to solve Eq.~\eqref{x_frac} for $\vartheta$, that is
\begin{align}
	\label{theta_of_x}
	\vartheta = \frac{x}{1-x}\,
	\frac{k_{\perp}^2}
	{Q^2 + P_{\perp}^2/
	\vartheta_{h1} 
	\vartheta_{h2}} 
	\sim
	\frac{ x\mkern1mu Q_s^2}
	{Q^2 + P_{\perp}^2/
	\vartheta_{h1} 
	\vartheta_{h2}},
\end{align}
where in writing the second, approximate, equality we have used the fact that the integration over $k_{\perp}$ is controlled by the regime $k_{\perp} \sim \tilde{Q}_s$ as already explained in Sect.~\ref{sec:dpdfs}. A direct inspection of this equality ensures that $\vartheta$ remains always small, as required by our approximations.

We can also define a ``typical'' hard jet momentum by setting $x_{\rm min} =1/2$. Then Eq.~\eqref{xmin} gives
\begin{align}
	\label{Pperp_typ}
	 P_{\perp\rm typ}^2  = 
	 \begin{cases}
	 	7.5\,{\rm GeV}^2 \quad &\mathrm{for} \quad \beta_{\min} = 0.2\phantom{1}
	 	\quad (\YP=4.5)
	 	\\*[0.2cm]
	 	18.3\,{\rm GeV}^2 \quad &\mathrm{for} \quad \beta_{\min} = 0.12
	 	\quad (\YP=4)
	 \end{cases}
\end{align}
and such hard momenta are indeed, as desired, well above the saturation momentum $Q_s^2$ and well below the corresponding maximum momenta in Eq.~\eqref{Pperp_max_numbers}. Taking into account all the above considerations, we shall present the results of our numerical calculations for $P_{\perp}^2$ varying from $2 Q_s^2$ up to a value close to $P_{\perp\max}^2$, keeping in mind that the accuracy may not be very good close to the end-points. On the one hand, when $P_{\perp}^2$ is not sufficiently large, saturation corrections of the order of $Q_s^2/P_{\perp}^2$ or higher (which have been neglected) get sizeable. On the other hand, when $P_{\perp}^2$ is too close to $P_{\perp\max}^2$ there is no phase space available for the production of the semi-hard jet.

Due to the definition of our observable, the CGC correlators $\mcal{W}_g$ and $\mcal{W}_q$ should be evaluated at a ``floating'' scale as we vary the hard jet momentum $P_{\perp}^2$ and/or the fraction $x$ (recall that $Q^2$, $\Ybj$ and the hard jet splitting fractions $\vartheta_{h1}$ and $\vartheta_{h2}$ are fixed), that is 
\begin{align}
	\label{YP_gen}
	\YP
	=
	Y_{_{\rm Bj}}
	-\ln\frac{Q^2 + P_{\perp}^2/\vartheta_{h1} 
	\vartheta_{h2}}{x\mkern1muQ^2},
\end{align}
where one should set $x=1$ for 2 jets. In principle this can be done although it will significantly complicate the numerical computation. We shall argue that it is not necessary to do so, by estimating the dependence on $P_{\perp}^2$ via the floating $\YP$ and showing that it is subdominant. The QCD correlators depend on $\YP$ via the ratio $\Delta_{\perp}^2/Q_s^2(\YP)$, thus one has
\begin{align}
	\label{deltaG}
	\Delta \mcal{W} \simeq 
	\frac{\dif \mcal{W}(\Delta_{\perp}^2/Q_s^2(\YP))}
	{\dif \YP}\,
	\Delta \YP
	\simeq 
	\frac{\dif \mcal{W}(\Delta_{\perp}^2/Q_s^2(\YP))}
	{\dif \ln Q_s^2(\YP)/\Delta_{\perp}^2}\,
	\frac{\dif \ln Q_s^2(\YP) }{\dif\YP}\,
	\Delta \YP.
\end{align}
The first factor in the above is of $\mcal{O}(1)$, since we are interested in a semi-hard momentum transfer $\Delta_{\perp}^2 \sim Q_s^2$. The second factor determines the growth of the saturation momentum, it is proportional to $\alpha_s$ and estimated to be around 0.25 \cite{GolecBiernat:1999qd,Triantafyllopoulos:2002nz}, while the last factor contains the $P_{\perp}^2$ dependence and can be easily determined from the explicit expression in Eq.~\eqref{YP_gen}. Focusing on the case of 2 jets just for simplicity and taking the variation with respect to the value at $\YP^{\min}$, we find
\begin{align}
	\label{deltaG2}
	\Delta \mcal{W} \sim \alpha_s 
	\ln \frac{Q^2 + P_{\perp\max}^2/
	\vartheta_{h1} \vartheta_{h2}}
	{Q^2 + P_{\perp}^2/
	\vartheta_{h1} \vartheta_{h2}}.
\end{align}
This kind of dependence of the cross sections on $P_{\perp}^2$ is mild since it is only logarithmic, unlike the strong power-law dependence through the hard factors. We will neglect it in our subsequent analysis, by computing the correlators at $\YP^{\min}$.

One should further realize that the QCD correlators are not the same for different values of $\YP^{\min}$. The most appropriate procedure, as explained in Sect.~\ref{sec:2jets}, would be to assume that the scattering amplitude is given by the MV model for a certain $Y_0$ and then evolve to the scale $\YP^{\min}$ according to the collinearly improved BK equation \cite{Beuf:2014uia,Iancu:2015vea,Ducloue:2019ezk}. Since we don't expect very significant changes in the form of the amplitude for an evolution between the two rapidities given in Eqs.~\eqref{mingap1} and \eqref{mingap2} which differ only by half a unit, and since the respective $\xP$'s are of the order of $10^{-2}$, we shall take the amplitude to be given by the MV model in both cases, but with different saturation scales. Fixing $Q_s^2 = 2\, {\rm GeV}^2$ at $\YP^{\min} = 4.5$, the aforementioned growth law of the saturation momentum (cf.~the discussion above Eq.~\eqref{deltaG2}) dictates us to set $Q_s^2 = 1.76\, {\rm GeV}^2$ for $\YP^{\min} = 4$.

Now we are ready to give the final form for the diffractive cross sections of interest. For 2 jets it is already given in Eq.~\eqref{sigmaave}, whereas for 2\,+\,1 jets we simply integrate Eqs.~\eqref{cs_soft_g} and \eqref{cs_soft_q} over the semi-hard jet kinematics to find
\begin{align}
	\hspace{0cm}
	\label{cs_mingap_g}
	\frac{\rmd\sigma^
	{\gamma_{\scriptscriptstyle \lambda}^* A\rightarrow q\bar q (g) X}}
	{\dif \Pi} = 
  	\alpha_{\rm em} \alpha_s
  	\sum\mkern-1.5mu e_f^2\,
  	\delta_{\vartheta}\,
  	H_\lambda^{(g)} (\vartheta_1,\vartheta_2,P_{\perp}, \bar{Q})
  	\int_{x_{\min}}^1\!
  	\frac{\dif x}{x}\,
  	\frac{\rmd xG_{\mathbb{P}}^{\rm inc}
  	(x, x_{\mathbb{P}}, P_{\perp}, \Delta_{\perp})}
  	{\rmd^2\bm{\Delta}},
\end{align}
and
\begin{align}
	\hspace{0cm}
	\label{cs_mingap_q}
	\frac{\rmd\sigma^
	{\gamma_{\scriptscriptstyle \lambda}^* A\rightarrow (q)\bar q g X}}
	{\dif \Pi} = 
  	2 \alpha_{\rm em} \alpha_s C_F
  	\sum\mkern-1.5mu e_f^2\,
  	\delta_{\vartheta}\,
  	H_\lambda^{(q)} (\vartheta_1,\vartheta_2,P_{\perp}, \tilde{Q})
  	\int_{x_{\min}}^1\!
  	\frac{\dif x}{x}\,
  	\frac{\rmd xq_{\mathbb{P}}^{\rm inc}
  	(x, x_{\mathbb{P}}, P_{\perp}, \Delta_{\perp})}
  	{\rmd^2\bm{\Delta}},
\end{align} 
with $x_{\min}$ defined in Eq.~\eqref{xmin} and where all cross sections should be used up to a maximum hard jet momentum which satisfies Eq.~\eqref{Pperp_max_numbers}. We will take the large-$N_c$ limit, i.e.~we will let $C_F \to N_c/2$ in Eqs.~\eqref{sigmaave} and \eqref{cs_mingap_q}, let $N_c^2-1 \to N_c^2$ in Eq.~\eqref{DPDF_def} for the gluon DPDF and use Eqs.~\eqref{appeq:Wg_gauss} and \eqref{appeq:Wq_gauss} for determining the scattering correlators $\mcal{W}_g$ and $\mcal{W}_q$. The QCD running coupling should be evaluated at the hard scale $P_{\perp}^2$ and at the one loop level we have
\begin{align}
	\label{RC}
	\alpha_s\big(P_{\perp}^2 \big) = 
	\frac{1}{b \ln P_{\perp}^2/\Lambda^2}
	\qquad \mathrm{with} \qquad
	b = \frac{11 N_c - 2 n_f}{12 \pi}.
\end{align}
Since it is equally probable to produce either a soft quark or a soft antiquark, the total 2\,+\,1 jets cross section is given by
\begin{align}
	\label{cs_mingap_tot}
	\frac{\rmd\sigma^
	{\gamma_{\scriptscriptstyle \lambda}^* A\rightarrow q\bar q g X}}
	{\dif \Pi}
	=
	\frac{\rmd\sigma^
	{\gamma_{\scriptscriptstyle \lambda}^* A\rightarrow q\bar q (g) X}}
	{\dif \Pi}
	+2\,
	\frac{\rmd\sigma^
	{\gamma_{\scriptscriptstyle \lambda}^* A\rightarrow (q)\bar q g X}}
	{\dif \Pi}. 
\end{align}

We can easily study the parametric dependence of the cross sections by using the explicit exact expressions for the hard factors in Eqs.~\eqref{hard_factor_2jets_T}, \eqref{hard_factor_2jets_L}, \eqref{HTgij_dec}, \eqref{HLg}, \eqref{HTQij} and \eqref{HLq} and the estimates for $\mcal{G}^{(+)}$ and the incoherent DPDFs in Eqs.~\eqref{Gplus_cases} and \eqref{DPDFs_qual} respectively. Assuming that $\bar{Q},\tilde{Q}\sim P_{\perp}$ and by neglecting some constant factors and logarithms we find
\begin{align}
	\label{cs_2jets_qual}
	\frac{\rmd\sigma^
	{\gamma_{\scriptscriptstyle \lambda}^* A\rightarrow q\bar q  X}}
	{\dif \Pi}
	\sim 
	\frac{S_{\perp} \alpha_{\rm em}}{4 \pi^3 N_c}
	\sum\mkern-1.5mu e_f^2\,
  	\delta_{\vartheta}\,
	\frac{Q_s^2}{P_{\perp}^6}	
	\times
	\begin{cases}
		\displaystyle \ln \frac{Q_s^2}{\Lambda^2}
	&\mathrm{for} \quad 
	\Delta_{\perp} \ll Q_s,
	 \\*[0.4cm]
 	\displaystyle\frac{Q_s^2}{\Delta_{\perp}^2}
 	&\mathrm{for} \quad
 	\Delta_{\perp} \gg Q_s.
	\end{cases}
\end{align}
for 2 jets and
\begin{align}
	\label{cs_3jets_qual}
	\frac{\rmd\sigma^
	{\gamma_{\scriptscriptstyle \lambda}^* A\rightarrow q\bar q g  X}}
	{\dif \Pi}
	\sim 
	\frac{S_{\perp} \alpha_{\rm em}}{4\pi^3 N_c}
	\sum\mkern-1.5mu e_f^2\,
  	\delta_{\vartheta}\,
	\frac{\alpha_s N_c}{\pi}\,
	\frac{1}{P_{\perp}^4}
	\times	
	\begin{cases}
		\displaystyle \ln \frac{Q_s^2}{\Lambda^2}
	&\mathrm{for} \quad 
	\Delta_{\perp} \ll Q_s,
	 \\*[0.4cm]
 	\displaystyle\frac{Q_s^4}{\Delta_{\perp}^4}
 	&\mathrm{for} \quad
 	\Delta_{\perp} \gg Q_s.
	\end{cases}
\end{align}
for 2\,+\,1 jets, where the soft gluon and soft quark sectors are of the same order. As expected, the above cross sections saturate logarithmically for $\Delta_{\perp} \ll Q_s$. The 2\,+\,1 jets cross section is parametrically larger not only when $\Delta_{\perp} \lesssim Q_s$ by a factor $\alpha_s \mkern1mu P_{\perp}^2 / Q_s^2$, but also when $Q_s^2 \ll \Delta_{\perp}^2 \ll P_{\perp}^2$ by a factor $\alpha_s \mkern1mu P_{\perp}^2/\Delta_{\perp}^2$. It is worth pointing out that such a dominant contribution exhibits a natural $1/\Delta_{\perp}^4 = 1/|t|^2$ tail at large $|t|$.   

\begin{figure}
	\begin{center}
	\includegraphics[width=0.48\textwidth]{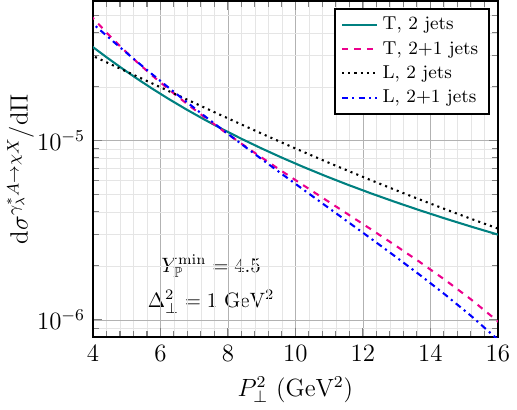}
	\hspace*{0.02\textwidth}
	\includegraphics[width=0.48\textwidth]{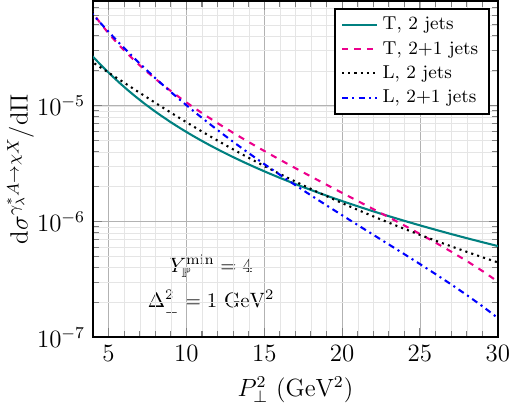}
	\end{center}
	\caption{\small The incoherent diffractive 2 jets and 2\,+\,1 jets cross sections for both transversely and longitudinally polarized virtual photons as functions of the hard jet momentum $P_{\perp}^2$ for fixed momentum transfer $\Delta_{\perp}^2$ and for $\YP^{\min}=4.5$ (left panel) or $\YP^{\min}=4$ (right panel). The remaining independent kinematic variables are fixed according to $\xbj=2.3 \times 10^{-3}$, $Q^2=20\mkern1.5mu{\rm GeV}^2$ and $\vartheta_{h1}=\vartheta_{h2}=1/2$. The scattering amplitude is given by the MV model with a saturation scale as described in the current section. A factor $\alpha_{\rm em} (S_{\perp}/4\pi^3) \delta_{\vartheta}$ has been neglected.}
\label{fig:Cross_section}
\end{figure}

We have computed numerically the cross sections in Eqs.~\eqref{sigmaave}, \eqref{cs_mingap_tot} and in Fig.~\ref{fig:Cross_section} we present the results as functions of $P_{\perp}^2$. The momentum transfer $\Delta_{\perp}^2$ is taken to be fixed and of the order of $Q_s^2$. The choice of the kinematic variables has been realistic and we are not exactly in the regime where the above parametric estimates are manifest. As a consequence, the 2\,+\,1 jets are not dominating the cross section. Still, we see that their contribution remains larger than the one from two jets up to a certain value of the hard jet momentum, which is roughly the typical one estimated in Eq.~\eqref{Pperp_typ}. Generally, the two contributions are more or less of the same order in the whole region in Fig.~\ref{fig:Cross_section}.

One may naively have expected a higher $P_{\perp}^2$ to give a more favorable situation for 2\,+\,1 jets. However, as we have already explained, a continuous increase in $P_{\perp}^2$ with a fixed $\YP^{\min}$ occurs at the expense of the phase space for the emission of the semi-hard jet. In the extreme case in which $P_{\perp}^2$ takes the maximal allowed value, cf.~\eqref{Pperp_max_numbers}, the cross section for 2\,+\,1 jets vanishes since there is no support in the $x$-integration in Eqs.~\eqref{cs_mingap_g} and \eqref{cs_mingap_q}.   

\begin{figure}
	\begin{center}
	\includegraphics[width=0.48\textwidth]{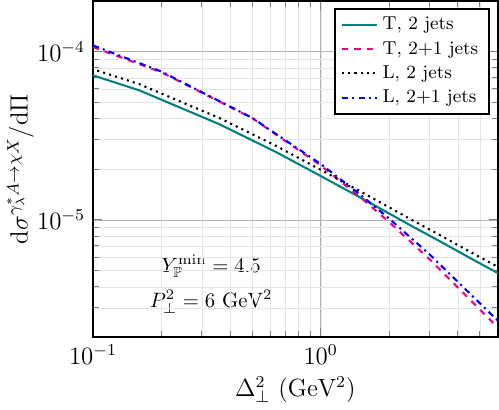}
	\hspace*{0.02\textwidth}
	\includegraphics[width=0.48\textwidth]{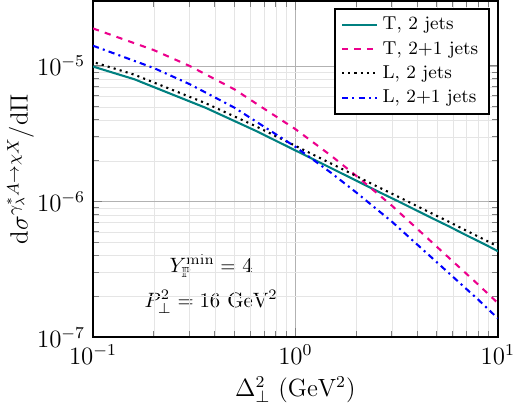}
	\end{center}
	\caption{\small The incoherent diffractive 2 jets and 2\,+\,1 jets cross sections for both transversely and longitudinally polarized virtual photons as functions of the momentum transfer $\Delta_{\perp}^2$ for fixed hard jet momentum $P_{\perp}^2$. The rest is as in Fig.~\ref{fig:Cross_section}.}
\label{fig:Cross_section_Delta}
\end{figure}

Similarly, in Fig.~\ref{fig:Cross_section_Delta} we show the numerically evaluated cross sections as functions of $\Delta_{\perp}^2$. The hard jet momentum $P_{\perp}^2$ is considered to be fixed and of the order of its typical value (which depends on $\YP^{\min}$), cf.~Eq.~\eqref{Pperp_typ}. We observe again that the two contributions are roughly of the same order in a wide range of momenta. With the chosen kinematics, the 2\,+\,1 jets cross section remains larger than the two jets one of for momentum transfers up the saturation scale. 

\section{Conclusion and perspectives}
\label{sec:conc}

In a previous work \cite{Rodriguez-Aguilar:2023ihz} we had studied in detail the incoherent diffractive production of (exactly) two jets in electron-nucleus DIS at small $\xbj$. For jets that have transverse momenta $P_{\perp}$ much larger than both the saturation scale $Q_s$ and the momentum transfer $\Delta_{\perp}$, so that they are hard and appear almost back-to-back in the transverse plane, we had found that the cross section falls rather fast with the hard jet momentum, namely like $1/P_{\perp}^6$, due to the fact that the scattering involves dipoles of small size. Moreover, when considering the dependence on the momentum transfer, we had obtained the somewhat strange $1/\Delta_{\perp}^2 = 1/|t|$ fall-off for large $|t|$. 

Here we showed, like for the corresponding problem in coherent diffraction \cite{Iancu:2021rup,Iancu:2022lcw,Iancu:2023lel,Hauksson:2024bvv}, that the leading contribution in power counting occurs when the hard dijet is accompanied by a third, semi-hard, jet. We wrote this 2\,+\,1 cross section in a factorized form in terms of a hard factor times an incoherent quark or gluon DTMD (if the semi-hard jet is explicit) or times the respective DPDF (if the semi-hard jet is integrated over). Although such a process is proportional to the running coupling $\alpha_s(P_{\perp}^2)$ which is small, the scattering is sensitive only to large size configurations in the projectile wave-function and as a result the hard factor of the 2\,+\,1 jets cross section falls only like $1/P_{\perp}^4$. Furthermore, we found that quark and gluon DPDFs saturate logarithmically when $|t| \ll Q_s^2$ and exhibit a natural $1/\Delta_{\perp}^4 = 1/|t|^2$ tail in the regime $Q_s^2 \ll |t| \ll P_{\perp}^2$. Finally, unlike for the two jets case \cite{Mantysaari:2019hkq,Rodriguez-Aguilar:2023ihz}, we didn't find a dependence of the 2\,+\,1 jets cross section on the angle between the hard jet momentum and the momentum transfer.

We have defined a suitable observable, namely the incoherent diffractive production of a hard dijet with a minimum rapidity gap, in order to make a more direct comparison between the 2 jets and 2\,+\,1 jets cross sections. The latter dominates not only parametrically, but also numerically at sufficiently high collision energies where we are allowed to explore very hard jet momenta for a given gap. On the other hand, considering some typical values of the kinematic variables, like those expected at the EIC, we found that the two cross sections are roughly of the same order. Thus, for current and near future practical purposes, we must take into account both contributions.

We have done only a ``proof of principle'' calculation by assuming a partonic content in the  outgoing state. Additional work is required in order to compute the cross sections for producing real jets or hadrons. In particular, it should be interesting and beneficial to consider $J/\psi$ production in $eA$ DIS or in UPCs of heavy nuclei \cite{Mantysaari:2022kdm} in the context of this work.

\begin{acknowledgments}
	D.N.T.~would like to thank Edmond Iancu for valuable discussions. S.Y.~Wei is supported by the Shandong Province Natural Science Foundation under grant No.~2023HWYQ-011 and the Taishan fellowship of Shandong Province for junior scientists.
\end{acknowledgments}

\appendix

\section{The gluon semi-hard tensor and its index structure}
\label{app:Gij}

The semi-hard tensor for incoherent diffraction can be obtained from the corresponding for the coherent case \cite{Iancu:2022lcw} by making the replacement
\begin{align}
	\label{coh_to_inc}
	\delta^{(2)}(\bm{\Delta})\mkern1mu
	\mcal{T}_g(R)\mkern1mu 
	\mcal{T}_g(\bar{R})
	\to 
	\int \frac{\dif^2 \bm{B}}{(2\pi)^2}\,
	e^{- i \bm{\Delta} \cdot \bm{B}}\,
	\mcal{W}_g
	(\bm{R},\bar{\bm{R}},\bm{B}).
\end{align}
We find
\begin{align}
	\label{Gij}
	\mcal{G}^{ij}_{\rm inc} (x,\xP,\bk_3,\bm{\Delta})=
	\frac{1}{4\pi}
	\int	\,&
	\frac{\rmd^2\bm{B}}{2\pi}\,
	\frac{\rmd^2\bm{R}}{2\pi}\,
	\frac{\rmd^2\bar{\bm{R}}}{2\pi}\,
	e^{- i \bm{\Delta} \cdot \bm{B}  - i \bk_3 \cdot (\bm{R}-\bar{\bm{R}})}\,
	\nonumber
	\\*[0.1cm]
	&\big(\hat{R}^{ik}
	\hat{\bar{R}}^{kj}+
	\hat{R}^{jk}
	\hat{\bar{R}}^{ki}
	\big)\,
	\mcal{M}^2 K_2(\mcal{M}R)\,
	\mcal{M}^2 K_2(\mcal{M}\bar{R})\,
	\mcal{W}_g
	(\bm{R},\bar{\bm{R}},\bm{B}),
\end{align}
where we have been allowed to symmetrize in the indices $i$ and $j$, by anticipating an eventual contraction with a symmetric hard tensor. Quite generally, we can now decompose the tensor $\mcal{G}^{ij}_{\rm inc}$ into a sum of a diagonal term and three traceless terms, namely
\begin{align}
	\label{Ginc_dec}
	\mcal{G}^{ij}_{\rm inc} = 
	\mcal{G}_{\rm inc}\,\frac{\delta^{ij}}{4}
	+ A_1\hat{\Delta}^{ij}
	+ A_2\mkern1mu\hat{k}_3^{ij}
	+ A_3\left(
	\frac{\Delta^i k_3^j}{\Delta_{\perp} k_{3\perp}} 
	+ \frac{\Delta^j k_3^i}{\Delta_{\perp} k_{3\perp}}  
	-\frac{\bm{\Delta} \mkern-1mu\cdot\mkern-1mu \bk_3}
	{\Delta_{\perp} k_{3\perp}}\,  
	\mkern1mu \delta^{ij}
	\right),
\end{align}
where $\mcal{G}_{\rm inc}$, $A_1$, $A_2$ and $A_3$ are scalars, that is, they depend only on $\Delta_{\perp}$, $k_{3\perp}$ and $\bm{\Delta} \mkern-2mu\cdot \mkern-2mu\bk_3$. It is straightforward to prove that
\begin{align}
	\label{RRV}
	\hat{R}^{ik}
	\hat{\bar{R}}^{kj}
	\hat{V}^{ij}=0
\end{align}
for three arbitrary tensors which are of the form given in Eq.~\eqref{Vij}, while the aforementioned symmetrization becomes important to show that
\begin{align}
	\label{RRVV}
	\big(\hat{R}^{ik}
	\hat{\bar{R}}^{kj}+
	\hat{R}^{jk}
	\hat{\bar{R}}^{ki}
	\big)
	\left(\frac{V_1^i V_2^j}{V_{1\perp}V_{2\perp}}
	-\frac{\bm{V}_1 \mkern-1mu\cdot\mkern-1mu \bm{V}_2}
	{V_{1\perp}V_{2\perp}}\,
	\frac{\delta^{ij}}{2}
	\right)
	=0
\end{align}
for arbitrary vectors $\bm{V}_1$ and $\bm{V}_2$. Contracting \eqref{Ginc_dec} with each one of the traceless tensors appearing in the decomposition and using the last two identities it is easy to show that $A_1=A_2=A_3=0$. Contracting with the diagonal $\delta^{ij}$ we arrive at the final form of the semi-hard gluon tensor given in Eqs.~\eqref{Gij_G} and \eqref{Ginc}.

\section{The incoherent gluon DPDF in limiting cases}
\label{app:gDPDF}

Here we will calculate analytically the dominant term of the incoherent gluon DPDF at low and high momenta in the limiting cases $x\ll 1$ and $1-x \ll 1$. We first rewrite Eq.~\eqref{DPDF_def}, that is 
\begin{align}
	\label{appeq:DPDF}
  \frac{\rmd xG_{\mathbb{P}}^{\rm inc}
  (x, x_{\mathbb{P}}, P_{\perp}, \Delta_{\perp})}
  {\rmd^2\bm{\Delta}}
  = \frac{S_{\perp} (N_c^2-1)}{4 \pi^3}\,
  \frac{1}{2\pi}
  \int \dif^2 \bk\,
  \frac{\mcal{G_{\rm inc}} (x, x_{\mathbb{P}}, \bk, \bm{\Delta})}
  {1-x},
\end{align}
where, as we have explained in Sect.~\ref{sec:dpdfs}, the integration over $\bk$ is unrestricted and thus the DPDF is $P_{\perp}$-independent. In the Gaussian approximation \cite{Iancu:2002aq,Dumitru:2011vk,Iancu:2011ns,Iancu:2011nj,Alvioli:2012ba} the multipoint correlator $\mcal{W}_g$, which enters in the definition of $\mcal{G}_{\rm inc}$ in Eq.~\eqref{Ginc}, reads \cite{Kovchegov:2012nd}
\begin{align}
	\label{appeq:Wg_gauss}
	\mcal{W}_g
	(\bm{R},\bar{\bm{R}},\bm{B}) =
	\frac{2}{N_c^2}\mkern1muf^2\mcal{S}_g(R)\, \mcal{S}_g(\bar{R})
	\left[ \frac{e^{-F}-1 +F}{F^2}+
	\frac{e^{-(F-f)}-1 +(F-f)}{(F-f)^2}
	\right],
\end{align}
where
\begin{align}
	\label{appeq:Fandf}
	F=\frac{1}{2} \ln \frac{
	\mcal{S}_g(R) 
	\mcal{S}_g(\bar{R})}
	{\mcal{S}_g(|\bm{B}+\bm{R} - \bar{\bm{R}}|) 
	\mcal{S}_g(B)}
	\qquad \mathrm{and} \qquad
	f=\frac{1}{2} \ln \frac{
	\mcal{S}_g(|\bm{B}+\bm{R}|) 
	\mcal{S}_g(|\bm{B}-\bar{\bm{R}}|)}
	{\mcal{S}_g(|\bm{B}+\bm{R} - \bar{\bm{R}}|) 
	\mcal{S}_g(B)}.
\end{align}
The above holds in the large-$N_c$ limit, i.e.~it does not take into account terms which are of the order of $1/N_c^4$ or higher, and is valid for arbitrary $\bm{R}$, $\bar{\bm{R}}$ and $\bm{B}$. At this point we open a parenthesis to give the corresponding expression for the correlator $\mcal{W}_q$ in the fundamental representation (which is used to numerically evaluate the incoherent quark DPDF), namely 
\begin{align}
	\label{appeq:Wq_gauss}
	\mcal{W}_q
	(\bm{R},\bar{\bm{R}},\bm{B}) =
	\frac{1}{N_c^2}\mkern1muf^2
	\mcal{S}(R)\mkern1mu 
	\mcal{S}(\bar{R})\,
	\frac{e^{-F}-1 +F}{F^2},
\end{align}
where we recall the Gaussian approximation relation $\mcal{S}(R) = [\mcal{S}_g(R)]^{C_F/N_c}$, which reduces to $\mcal{S}(R) \simeq \sqrt{\mcal{S}_g(R)}$ at large-$N_c$. We will also take $\mcal{S}_g(R)$ to be given by the MV model, cf.~Eqs.~\eqref{SgMV} and \eqref{QAQs}.

It is worthwhile to observe the similar functional form of the expressions in Eqs.~\eqref{appeq:Wg_gauss} and \eqref{appeq:Wq_gauss}. The main difference lies in the presence of the two terms in Eq.~\eqref{appeq:Wg_gauss} which make the square bracket symmetric under the exchange $F \to F-f$. This ensures that the gluon correlator $\mcal{W}_g$ remains invariant if we swap the positions of the gluons in one of the two dipoles.

\subsection{The case $x\ll 1$}
\label{app:smallx}

When $x \ll 1$, we have $\mcal{M} \simeq k_{\perp}\sqrt{x} \ll k_\perp$ and the contribution from the regime $k_{\perp} R \gtrsim 1/\sqrt{x} \gg 1$ in which the argument of the $K_2$ Bessel functions becomes of the order of (or larger than) one is suppressed, due to the strong oscillations induced by $e^{-i \bk \cdot \bm{R}}$. Therefore we are allowed to expand the Bessel functions for small argument and using $\mcal{M}^2 K_2(\mcal{M}R) \simeq 2/R^2$ (and similarly for $\mcal{M}^2 K_2(\mcal{M}\bar{R})$) we see that the integration over $\bk$ becomes trivial and leads to $(2\pi)^2\delta^{(2)}(\bm{R} - \bar{\bm{R}})$. This further allows us to integrate directly over $\bar{\bm{R}}$ and we obtain the rather simple form
\begin{align}
	\label{appeq:int_Ginc_0}
	\int \dif^2\bk\,
	\mcal{G}_{\rm inc}=
	4 \int	
	\frac{\rmd^2\bm{B}}{2\pi}\,
	e^{- i \bm{\Delta} \cdot \bm{B}}
	\int 
	\frac{\rmd^2\bm{R}}{2\pi}\,
	\frac{\mcal{W}_g
	(\bm{R},\bar{\bm{R}}=\bm{R},\bm{B})}{R^4}.
\end{align}
Since we set $\bar{\bm{R}} = \bm{R}$, the functions $F$ and $f$ in Eq.~\eqref{appeq:Fandf} simplify and read
\begin{align}
	\label{appeq:Fandf_R}
	F=\ln \frac{\mcal{S}_g(R)}
	{\mcal{S}_g(B)}
	\qquad \mathrm{and} \qquad
	f=\frac{1}{2} \ln \frac{\mcal{S}_g(|\bm{B}+\bm{R}|) 
	\mcal{S}_g(|\bm{B}-\bm{R}|)}
	{\mcal{S}_g^2(B)}.
\end{align}

\paragraph{Low momenta $\Delta_{\perp} \ll Q_s$.}
The dominant contribution in this regime is coming from configurations such that $B \gg 2/Q_s, R$, but we do not assume any special ordering between $R$ and $2/Q_s$. We readily find
\begin{align}
	\label{appeq:F0low}
	F \simeq F-f \simeq -\ln \mcal{S}_g(B) = 
	\frac{Q_A^2 B^2}{4}\,
	\ln \frac{4}{B^2 \Lambda^2} \gg 1.
\end{align}
When considering $f$ the leading term cancels and performing a Taylor expansion for $R \ll B$, we find the dominant, logarithmically enhanced, term\footnote{One can set $R$ equal to zero when it appears next to $B$ in the argument of a logarithm. We shall repeatedly employ this property and the analogous one for $B \ll R$ needed in the high momentum region, throughout this Appendix.}
\begin{align}
	\label{appeq:f0low}
	f \simeq 
	-\frac{Q_A^2 R^2}{4} \ln \frac{4}{B^2 \Lambda^2}.
\end{align}
Inserting the last two equations into Eq.~\eqref{appeq:Wg_gauss}, we see that the correlator becomes
\begin{align}
	\label{appeq:w0low}
	\mcal{W}_g(\bm{R},\bm{R},\bm{B})
	\simeq
	\frac{1}{N_c^2}\,Q_A^2 R^4 \mcal{S}_g^2(R)\,
	\frac{\ln4/B^2\Lambda^2}{B^2}.
\end{align}
The integration over $\bm{R}$ is controlled by dipole sizes of the order of $2/Q_s$. More precisely, we easily find in the MV model\footnote{The leading result is the same as the one obtained using the GBW model. Employing the MV model one can calculate corrections suppressed by inverse powers of $\ln Q_s^2/\Lambda^2$ \!\cite{Iancu:2004bx} which we neglect here.} 
\begin{align}
\label{appeq:intS2}
	\int_0^\infty \dif R\, R\, 
	\mcal{S}_g^2(R)
	\simeq
	\frac{1}{Q_s^2},
\end{align}
where we have been allowed to extend the integration to infinity (since, as we already pointed out, $R$ is not restricted w.r.t.~to $2/Q_s$) and thus we have also been able to keep track of the exact coefficient. Recalling that $Q_s^2 = Q_A^2 \ln Q_s^2/\Lambda^2$, we get 
\begin{align}
	\label{appeq:Ginc_0low}
	\int \dif^2\bk\,
	\mcal{G}_{\rm inc} \simeq 
	\frac{4}{N_c^2}\,
	\frac{1}{\ln Q_s^2/\Lambda^2}
	\int	
	\frac{\rmd^2\bm{B}}{2\pi}\,
	e^{- i \bm{\Delta} \cdot \bm{B}}\, 
	\frac{\ln4/B^2\Lambda^2}{B^2}.
\end{align}
After performing the trivial angular integration, we observe that the integration becomes logarithmic in the regime $2/Q_s \ll  B \ll 2/\Delta_{\perp}$. By further taking the large-$N_c$ limit of the factor multiplying the integration in Eq.~\eqref{appeq:DPDF}, we finally arrive at
\begin{align}
\label{appeq:DPDF_0low}
  \frac{\rmd xG_{\mathbb{P}}^{\rm inc}}
  {\rmd^2\bm{\Delta}}
  \simeq
  \frac{S_{\perp}}{8 \pi^4}\,
  \frac{1}{\ln Q_s^2/\Lambda^2}
	\left(
	\ln^2 \frac{Q_s^2}{\Lambda^2} - 
	\ln^2 \frac{\Delta_{\perp}^2}{\Lambda^2} 
	\right)	
	\quad \mathrm{for} \quad
	x \ll 1 
	\quad \mathrm{and} \quad
	\Delta_{\perp} \ll Q_s.
\end{align}

\paragraph{High momenta $\Delta_{\perp} \gg Q_s$.} Now the dominant contribution is expected to come from dipoles with a small size $B$, more precisely we shall see that there is a logarithmic enhancement in the regime $ B \ll R \ll 2/Q_s$. Since $\mcal{S}(R) \simeq 1$ and $F,f,F-f \ll 1$, we immediately see that the correlator becomes $\mcal{W}_g \simeq 2 f^2/N_c^2$ in the 4-gluon exchange approximation. Expanding $f$ in the regime $B\ll R$ we find
\begin{align}
	\label{appeq:f0high}
	\mcal{W}_g(\bm{R},\bm{R},\bm{B}) 
	\simeq
	\frac{Q_A^4 R^4}{8 N_c^2}\ln^2\frac{4}{R^2 \Lambda^2} 
	+\frac{Q_A^4 B^2 R^2}{4 N_c^2}
	\left(\ln^2 \frac{4}{R^2 \Lambda^2}
	- \ln \frac{4}{B^2 \Lambda^2} \ln\frac{4}{R^2 \Lambda^2}
	\right).
\end{align}
The first term in the above is independent of $\bm{B}$ and is not relevant to our purposes since after the Fourier transform, cf.~\eqref{appeq:DPDF}, it will become proportional to $\delta^{(2)}(\bm{\Delta})$.\footnote{Such a term is important for the integrated over $\bm{\Delta}$ cross section, however one must relax the assumption $R \ll 2/Q_s$ in order to do properly the corresponding calculation.} Inserting the second term of Eq.~\eqref{appeq:f0high} into Eq.~\eqref{appeq:int_Ginc_0} we see that the integration over $R$ becomes logarithmic (recall that $ B \ll R \ll 2/Q_s$) and gives
\begin{align}
	\label{appeq:Ginc_0high}
	\int \dif^2\bk\,
	\mcal{G}_{\rm inc} 
	\simeq
	-\frac{Q_A^4}{12 N_c^2}
	\int	
	\frac{\rmd^2\bm{B}}{2\pi}\,
	e^{- i \bm{\Delta} \cdot \bm{B}}\, 
	B^2
	\left(
	\ln^3 \frac{4}{B^2 Q_s^2} + 
	3 \ln \frac{Q_s^2}{\Lambda^2}\,
	\ln^2 \frac{4}{B^2 Q_s^2}
	\right).
\end{align}
Since the above is proportional to $B^2$, it becomes clear from Eq.~\eqref{appeq:DPDF_0low} that the incoherent DPDF must scale like $1/\Delta_{\perp}^4$. Still, we point out that the presence of the logarithmic dependence on $B$ in Eq.~\eqref{appeq:Ginc_0high} is necessary\footnote{These logarithms have their origin in the logarithm of the MV model amplitude. Had we used the GBW model, we would not have found any power-law tail at large $\Delta_{\perp}$. The result would have been an unphysical distribution which decays exponentially for large $\Delta_{\perp}$.} in order to get a non-zero result for $\Delta_{\perp} \neq 0$, since
\begin{align}
	\label{appeq:FT_B2}
	\int \frac{\rmd^2\bm{B}}{2\pi}\,
	e^{- i \bm{\Delta} \cdot \bm{B}}
	B^2 = -\nabla^2_{\Delta}
	\int \frac{\rmd^2\bm{B}}{2\pi}\,
	e^{- i \bm{\Delta} \cdot \bm{B}} = -2 \pi \nabla^2_{\Delta} \delta^{(2)}(\bm{\Delta}).
\end{align}
Doing the angular integration, letting $B = \tilde{B}/\Delta_{\perp}$ and using
 \begin{align}
 	\label{appeq:FT_B2_log}
 	\int_0^{\infty} \dif \tilde{B}\, \tilde{B}^3 J_0(\tilde{B}) \ln \tilde{B}^2 =8, 
 \end{align}
we finally obtain the leading behavior
\begin{align}
\label{appeq:DPDF_0high}
  \frac{\rmd xG_{\mathbb{P}}^{\rm inc}}
  {\rmd^2\bm{\Delta}}
  \simeq
  \frac{S_{\perp}}{4 \pi^4}\,
  \frac{Q_A^4}{\Delta_{\perp}^4}
	\left( 
	\ln^2 \frac{\Delta_{\perp}^2}{\Lambda^2} -
	\ln^2 \frac{Q_s^2}{\Lambda^2} 
	\right)	
		\quad \mathrm{for} \quad
	x \ll 1 
	\quad \mathrm{and} \quad
	\Delta_{\perp} \gg Q_s.
\end{align}

\subsection{The case $1- x \ll 1$}
\label{app:largex} 

When $1-x \ll 1$, we have $\mcal{M} \simeq k_{\perp}/\sqrt{1-x} \gg k_\perp$ and the contribution from the regime $k_{\perp} R \gtrsim 1$ is suppressed, due to the strong exponential decay of the $K_2$ Bessel functions. Therefore we are allowed to set $e^{- i \bk \cdot (\bm{R} - \bar{\bm{R}})} \simeq 1$ and the integration over $\bk$ can be done analytically to give
\begin{align}
	\label{appeq:int_Ginc_1}
	\int \dif^2\bk\,
	\frac{\mcal{G}_{\rm inc}}{1-x}\bigg|_{x=1}=
	\int \frac{\rmd^2\bm{B}}{2\pi}\,
	e^{- i \bm{\Delta} \cdot \bm{B}}\!
	\int \frac{\rmd^2\bm{R}}{2\pi}\,
	\int \frac{\rmd^2\bar{\bm{R}}}{2\pi}\,
	\frac{\Psi(\bar{R}/R)}{R^3 \bar{R}^3}\,
	\cos 2 \phi_{R\bar{R}}\,
	\mcal{W}_g
	(\bm{R},\bar{\bm{R}},\bm{B}),
\end{align}
where the dimensionless function $\Psi$ reads
\begin{align}
	\label{appeq:Psirho}
	\Psi(\rho) =  
	\frac{16 \rho}{(\rho^2-1)^5}\,
	\left(
	\rho^8 - 8 \rho^6 + 8 \rho^2 -1 +12 \rho^4\ln \rho^2
	\right).    
\end{align} 
We will find convenient to make a change of variables from the size $\bar{R}$ to the ratio $\rho \equiv \bar{R}/R$. By definition, the function in Eq.~\eqref{appeq:Psirho} satisfies $\Psi(1/\rho) = \Psi(\rho)$, since the starting integral is symmetric under the exchange $R \leftrightarrow \bar{R}$. Due to this symmetry, we will integrate only for $\bar{R}<R \Leftrightarrow \rho<1$ and multiply by a factor of 2. It will also  be helpful to know that $\Psi(\rho) \simeq 16 \rho$ for  $\rho \ll 1$, while the function is finite at $\rho=1$, namely $\Psi(1) = 32/5$. 

\paragraph{Low momenta $\Delta_{\perp} \ll Q_s$.} The dominant configurations are those with sizes that satisfy $B \gg 2/Q_s, R,\bar{R}$, but other than that there isn't any special constraint among $2/Q_s$,\,$R$ and $\bar{R}$. Concerning $F$ and $F-f$, they are still given by Eq.~\eqref{appeq:F0low}, since only the largest size $B$ is relevant for their determination. The leading term in the expansion of $f$ for $R, \bar{R} \ll B$ cancels and we obtain 
\begin{align}
	\label{appeq:f1low}
	f \simeq 
	-\frac{Q_A^2 \bm{R} \cdot \! \bar{\bm{R}}}{4} \ln \frac{4}{B^2 \Lambda^2} = 
	-\frac{Q_A^2 R^2}{4}\rho \cos \phi_{R\bar{R}} \ln \frac{4}{B^2 \Lambda^2}.
\end{align}
Then Eq.~\eqref{appeq:Wg_gauss} gives
\begin{align}
	\label{appeq:w1low}
	\mcal{W}_g(\bm{R},\bar{\bm{R}},\bm{B})
	\simeq
	\frac{1}{N_c^2}\, 
	Q_A^2 R^4 
	\mcal{S}_g(R) \mkern1mu
	\mcal{S}_g(\rho R) \mkern1mu
	\rho^2 \cos^2\mkern-2mu \phi_{R\bar{R}}\,
	\frac{\ln4/B^2\Lambda^2}{B^2}
\end{align}
and it becomes trivial to perform the two angular integrations (using, for example, $\phi_{R\bar{R}}$ and the angle between $\bm{B}$ and $\bm{R}$ as independent variables) to find
\begin{align}
	\label{appeq:Ginc_1low}
	\int \dif^2\bk\,
	\frac{\mcal{G}_{\rm inc}}{1-x}\bigg|_{x=1}
	\simeq
	\frac{Q_A^2}{4N_c^2}
	\int \frac{\rmd^2\bm{B}}{2\pi}\,
	e^{- i \bm{\Delta} \cdot \bm{B}}\,
	\frac{\ln4/B^2\Lambda^2}{B^2}
	\int_0^1\! \rmd \rho\, \Psi(\rho)
	\int_0^{\infty}\! \rmd R^2 \mkern1mu
	\mcal{S}_g(R) 
	\mkern1mu
	\mcal{S}_g(\rho R).
\end{align}
The remaining integrations over $R$ and $\rho$ are also straightforward, namely
\begin{align}
\label{appeq:intSS_Psi1}
	\int_0^{\infty}\! \rmd R^2 
	\mcal{S}_g(R) \mkern1mu 
	\mcal{S}_g(\rho R) \mkern1mu
	\simeq
	\frac{4}{(1+\rho^2)Q_s^2}
\qquad \mathrm{and} \qquad
	\int_0^1\! \rmd \rho\, 
	\frac{\Psi(\rho)}{1+\rho^2} = \frac{3\pi^2 -16}{4},
\end{align}
so that Eq.~\eqref{appeq:Ginc_1low} becomes
\begin{align}
	\label{appeq:Ginc_1low_b}
	\int \dif^2\bk\,
	\frac{\mcal{G}_{\rm inc}}{1-x}\bigg|_{x=1}
	\simeq
	\frac{3\pi^2 -16}{4N_c^2}
	\frac{1}{\ln Q_s^2/\Lambda^2}
	\int	
	\frac{\rmd^2\bm{B}}{2\pi}\,
	e^{- i \bm{\Delta} \cdot \bm{B}}\, 
	\frac{\ln4/B^2\Lambda^2}{B^2},
\end{align}
where we have used once more $Q_s^2 = Q_A^2 \ln Q_s^2/\Lambda^2$. The last integral is identical to the one appearing in Eq.~\eqref{appeq:Ginc_0low}. By further taking the large-$N_c$ limit of the factor multiplying the integration in Eq.~\eqref{appeq:DPDF}, we finally arrive at
\begin{align}
\label{appeq:DPDF_1low}
\hspace{-0.4cm}
  \frac{\rmd xG_{\mathbb{P}}^{\rm inc}}
  {\rmd^2\bm{\Delta}}
  \simeq
  \frac{S_{\perp}}{8 \pi^4}\,
  \frac{3\pi^2 - 16}{16}\,
  \frac{1}{\ln Q_s^2/\Lambda^2}
	\left(
	\ln^2 \frac{Q_s^2}{\Lambda^2} - 
	\ln^2 \frac{\Delta_{\perp}^2}{\Lambda^2} 
	\right)	
	\quad \mathrm{for} \quad
	1-x \ll 1 
	\quad \mathrm{and} \quad
	\Delta_{\perp} \ll Q_s.
\end{align}
We note that the above and the respective expression for $x \ll 1$ in Eq.~\eqref{appeq:DPDF_0low} have exactly the same functional form and that the numerical values of the two coefficients are very close to each other.

\paragraph{High momenta $\Delta_{\perp} \gg Q_s$.} This is the most complicated calculation of the Appendix and we will shall give a concise presentation by showing the essential steps. The relevant configurations are those which satisfy the strong ordering $B \ll R,\bar{R} \ll 2/Q_s$, but with no constraint between $R$ and $\bar{R}$. The correlator in Eq.~\eqref{appeq:Wg_gauss} reduces to $\mcal{W}_g \simeq 2 f^2/N_c^2$, with
\begin{align}
	\label{appeq:f1high}
	f \to -\frac{Q_A^2}{8}
	\bigg\{
	& R^2 \ln \frac{4}{R^2\Lambda^2}
	+\bar{R}^2 \ln \frac{4}{\bar{R}^2\Lambda^2}
	-(\bm{R} \!-\! \bar{\bm{R}})^2 
	\ln \frac{4}{(\bm{R} \!-\! \bar{\bm{R}})^2 \Lambda^2}
	\nn
	&+B^2\left[
	 \ln \frac{4}{R^2\Lambda^2}
	 +\ln \frac{4}{\bar{R}^2\Lambda^2}
	 -\ln \frac{4}{B^2\Lambda^2}
	 - \ln \frac{4}{(\bm{R} \!-\! \bar{\bm{R}})^2 \Lambda^2}
	\right]
	\bigg\}.
\end{align}
Even though $f$ contains a linear in $\bm{B}$ term, we have not included it since it can be shown that it does not contribute. We are clearly interested only in the term of $\mcal{W}_g$ which is proportional to $B^2$, therefore only the ``cross-term'' matters when squaring the r.h.s.~in Eq.~\eqref{appeq:f1high}. After some straightforward algebra we find
\begin{align}
\hspace{-0.5cm}
	\label{appeq:w1high}
	\mcal{W}_g(\bm{R},\bar{\bm{R}},\bm{B})
	\to \frac{Q_A^4 B^2 R^2}{16 N_c^2}
	&\left[ 
	2 \rho \cos\phi_{R\bar{R}}\, 
	\ln \frac{4}{R^2\Lambda^2}
	+\big(1 - 2 \rho \cos\phi_{R\bar{R}}+\rho^2\big)
	\ln\! \big(1 - 2 \rho \cos\phi_{R\bar{R}}+\rho^2\big)
	\right]
	\nn
	&\,\,\,\times 
	\left[
	\ln \frac{B^2}{R^2} +
	\ln\! \big(1 - 2 \rho \cos\phi_{R\bar{R}}+\rho^2\big)
	\right],
\end{align}
where we have dropped terms which involve $\ln \rho^2$ since the regime $\rho \ll 1$ does not lead to any logarithmic enhancement. Only logarithms which involve two of the very disparate scales $B$,\,$R$ and $2/\Lambda$ are large. Moreover, when multiplying the two brackets in Eq.~\eqref{appeq:w1high}, only the terms which are linear in $\ln\! \big(1 - 2 \rho \cos\phi_{R\bar{R}}+\rho^2\big)$ are relevant. Indeed, the term without such a logarithm vanishes after integrating over the angles, whereas the quadratic one is not logarithmically large. Thus Eq.~\eqref{appeq:w1high} simplifies to
\begin{align}
\hspace{-0.3cm}
	\label{appeq:w1high_b}
	\mcal{W}_g(\bm{R},\bar{\bm{R}},\bm{B}) 
	\to \frac{Q_A^4 B^2 R^2}{16 N_c^2}
	\left[ 
	2 \rho \cos\phi_{R\bar{R}}\, 
	\ln \frac{4}{B^2\Lambda^2}
	+\big(1 + \rho^2\big)
	\ln \frac{B^2}{R^2}
	\right]
	\ln\! \big(1 - 2 \rho \cos\phi_{R\bar{R}}+\rho^2\big).
\end{align}
Inserting the above into Eq.~\eqref{appeq:int_Ginc_1} and doing the angular integrations we readily obtain
\begin{align}
	\label{appeq:Ginc_1high}
	\int \dif^2\bk\,
	\frac{\mcal{G}_{\rm inc}}{1-x}\bigg|_{x=1}
	\simeq\,\,  &
	\frac{Q_A^4}{32 N_c^2}
	\int \frac{\rmd^2\bm{B}}{2\pi}\,
	e^{- i \bm{\Delta} \cdot \bm{B}}
	B^2
	\nn[0.1cm]
	&\times \int_{B^2}^{4/Q_s^2} \frac{\dif R^2}{R^2}
	\int_0^1 \dif \rho\, \Psi(\rho)
	\left[
	 (1+\rho^2)\ln\frac{R^2}{B^2}
	-\frac{2}{3} (3+\rho^2) \ln\frac{4}{B^2\Lambda^2}
	\right].
\end{align}
In principle the lower limit in the integration over $\rho$ is $B/R \ll 1$, but it has safely been set equal to zero, since the integrand is proportional to $\rho$ in this regime. Such an integration is straightforward and leads to 
\begin{align}
	\label{appeq:Ginc_1high_b}
	\int \dif^2\bk\,
	\frac{\mcal{G}_{\rm inc}}{1-x}\bigg|_{x=1}
	\simeq 
	\frac{Q_A^4}{24 N_c^2}
	\int \frac{\rmd^2\bm{B}}{2\pi}\,
	e^{- i \bm{\Delta} \cdot \bm{B}}
	B^2
	\int_{B^2}^{4/Q_s^2} \frac{\dif R^2}{R^2}
	\left(
	5\ln\frac{R^2}{B^2}
	-
	8 \ln\frac{4}{B^2\Lambda^2}
	\right).
\end{align}
Next we perform the logarithmic integration over $R^2$ to obtain
\begin{align}
	\label{appeq:Ginc_1high_c}
	\int \dif^2\bk\,
	\frac{\mcal{G}_{\rm inc}}{1-x}\bigg|_{x=1}
	\simeq 
	-\frac{Q_A^4}{48 N_c^2}
	\int \frac{\rmd^2\bm{B}}{2\pi}\,
	e^{- i \bm{\Delta} \cdot \bm{B}}
	B^2
	\left(11 \ln^2\frac{4}{B^2 Q_s^2}
	+ 16 \ln \frac{Q_s^2}{\Lambda^2}
	\ln\frac{4}{B^2Q_s^2}
	\right).
\end{align}
Finally, following the steps as in going from Eq.~\eqref{appeq:Ginc_0high} to Eq.~\eqref{appeq:DPDF_0high}, we arrive at
\begin{align}
	\label{appeq:DPDF_1high}
	\frac{\rmd xG_{\mathbb{P}}^{\rm inc}}
  	{\rmd^2\bm{\Delta}}	
  	\simeq
  	\frac{S_{\perp}}{8 \pi^4}\,
	\frac{Q_A^4}{3\Delta_{\perp}^4}
	\left(11 
	\ln \frac{\Delta_{\perp}^2}{\Lambda^2} -
	3 \ln \frac{Q_s^2}{\Lambda^2} 
	\right)	
	\quad \mathrm{for} \quad
	1-x \ll 1 
	\quad \mathrm{and} \quad
	\Delta_{\perp} \gg Q_s.
\end{align}

Before closing this Appendix, let us elaborate a bit more on the $x$-dependence when $x$ is close to one. One way to integrate over the gluon transverse momentum (which leads to Eq.~\eqref{appeq:int_Ginc_1}), is to make a change of variable from $k_{\perp}$ to $\mcal{M}$ by using the relation $k_{\perp}^2 \simeq (1-x) \mcal{M}^2$. Since $\mcal{M}$ is the scale which enters the argument of the $K_2$ Bessel functions, and since $e^{- i \bk \cdot (\bm{R} - \bar{\bm{R}})} \simeq 1$, it becomes clear that the only dependence on $1-x$ is the one from the change in the measure, i.e.~from $\dif k_{\perp}^2 \simeq (1-x)\mkern1mu \dif\mcal{M}^2$. This $1-x$ factor cancels the corresponding one appearing the denominator in Eq.~\eqref{DPDF_def}, so that eventually the incoherent DPDF is independent of $1-x$. However, it is extremely important to notice that the integrand in Eq.~\eqref{appeq:int_Ginc_1} contains an oscillating factor $\cos 2 \phi_{R\bar{R}}$. The reason for which we obtain a non-vanishing result when performing the angular integration, is that there is an additional multiplicative dependence on $\cos 2 \phi_{R\bar{R}}$ (or equivalently on $\cos^2\mkern-2mu\phi_{R\bar{R}}$) from the CGC scattering correlator $\mcal{W}_g(\bm{R},\bar{\bm{R}},\bm{B})$, cf.~Eqs.~\eqref{appeq:w1low} and \eqref{appeq:w1high_b}. This is very different from what happens in the case of the coherent DPDF, where the corresponding correlator $\mcal{T}_g(R)\mkern1mu \mcal{T}_g(\bar{R})$ is angle independent. In order to have a non-zero result, one must expand the exponential $e^{- i \bk \cdot (\bm{R} - \bar{\bm{R}})}$ to the fourth order to generate the desired angular dependence \cite{Iancu:2022lcw}. When we make the aforementioned change of variable to $\mcal{M}^2$, the extra power $k_{\perp}^4$ leads to a multiplicative factor $(1-x)^2$ which is transmitted to the coherent gluon DPDF. Similarly, the integrand of the quark DPDF involves $\cos \phi_{R\bar{R}}$, thus it is enough to expand $e^{- i \bk \cdot (\bm{R} - \bar{\bm{R}})}$ only to the second order and then the extra power $k_{\perp}^2$ leads to a multiplicative factor $1-x$ \cite{Hauksson:2024bvv}.


\providecommand{\href}[2]{#2}\begingroup\raggedright\endgroup

\end{document}